\documentclass[12pt,a4paper,final]{article}
\usepackage[utf8]{inputenc}
\usepackage[english]{babel}
\usepackage{amsmath}
\usepackage{amsfonts}
\usepackage{amssymb}
\usepackage{appendix}
\usepackage{faktor}
\usepackage[hidelinks]{hyperref}
\hypersetup{
    colorlinks=true,
    urlcolor=blue,
    linkcolor=blue,
    citecolor=red,
    }
\usepackage{amsthm}
\usepackage{bbold}
\usepackage{mathrsfs}
\usepackage{wasysym}
\usepackage{mathtools}
\usepackage{bm}
\usepackage{cancel}
\usepackage{color}
\usepackage{tikz-cd}
\usepackage[Glenn]{fncychap}
\usepackage{subfig}
\usepackage{accents}
\usepackage{slashed}
\usepackage{psfrag}
\usepackage{stackengine}
	\newcommand\ubar[1]{\stackunder[1.2pt]{$#1$}{\rule{.8ex}{.1ex}}}
\usetikzlibrary{babel}
\makeatletter
\newcommand*\owedge{\mathpalette\@owedge\relax}
\newcommand*\@owedge[1]{%
	\mathbin{%
		\ooalign{%
			$#1\m@th\bigcirc$\cr
			\hidewidth$#1\m@th\wedge$\hidewidth\cr
		}%
	}%
}
\makeatother

\newtheorem{teo}{Theorem}[section]
\newtheorem{cor}[teo]{Corollary}
\newtheorem{prop}[teo]{Proposition}
\newtheorem{lema}[teo]{Lemma}
\newtheorem{defi}[teo]{Definition}

\newtheorem{rmk}[teo]{Remark}

\usepackage{makeidx}
\usepackage{graphicx}
\usepackage[left=2.60cm, right=2.60cm, top=2.45cm, bottom=2.80cm]{geometry}

\usepackage{scalerel}
\usepackage{stackengine,wasysym}

\usepackage{color}
    \makeatletter
\renewcommand\part{%
	\if@openright
	\cleardoublepage
	\else
	\clearpage
	\fi
	\thispagestyle{empty}%
	\if@twocolumn
	\onecolumn
	\@tempswatrue
	\else
	\@tempswafalse
	\fi
	\null\vfil
	\secdef\@part\@spart}
\makeatother
\newcommand{\II}{\mathrm {I\!I}}

\newcommand{\wh}{\widehat}

\newcommand{\ul}{\underline}
\newcommand{\ol}{\overline}

\newcommand{\rad}{\operatorname{Rad}}

\newcommand{\bY}{\textup{\textbf Y}}

\newcommand{\bF}{\textup{\textbf F}}
\newcommand{\bK}{\textup{\textbf K}}
\newcommand{\bPi}{\bm\Pi}
\newcommand{\bg}{\bm{\gamma}}
\newcommand{\bt}{\bm{\tau}}

\newcommand{\mf}{\mathfrak}
\newcommand{\real}{\mathbb R}
\newcommand{\tr}{\operatorname{tr}}

\newcommand{\grad}{\text{grad}}

\newcommand{\st}{\stackrel}

\newcommand{\para}{\parallel}

\newcommand{\X}{\mathfrak{X}}
\renewcommand{\d}{\coloneqq}

\newcommand{\ric}{\operatorname{Ric}}

\newcommand{\lie}{\mathcal{L}}
\newcommand{\mc}{\mathcal}

\renewcommand{\to}{\longrightarrow}

\renewcommand{\mapsto}{\longmapsto}
\title{\vspace*{-1.35cm}\textbf{Covariant definition of Double Null Data and geometric uniqueness of the characteristic initial value problem}}
\author{Marc Mars\footnote{\href{mailto:marc@usal.es}{marc@usal.es}}\,\, and Gabriel Sánchez-Pérez\footnote{\href{mailto:gasape21@usal.es}{gasape21@usal.es}} \\
	Departamento de Física Fundamental, Universidad de Salamanca\\
	Plaza de la Merced s/n, 37008 Salamanca, Spain }
\date{\today}

\begin{document}
	\maketitle
\vspace{-2mm}
\begin{abstract}
The characteristic Cauchy problem of the Einstein field equations has been recently addressed from a completely abstract viewpoint by means of hypersurface data and, in particular, via the notion of double null data. However, this definition was given in a partially gauge-fixed form. In this paper we generalize the notion of double null data in a fully diffeomorphism and gauge covariant way, and show that the definition is complete by proving that no extra conditions are needed to embed the double null data in some spacetime. The second aim of the paper is to show that the characteristic Cauchy problem satisfies a geometric uniqueness property. Specifically, we introduce a natural notion of isometry at the abstract level such that two double null data that are isometric in this sense give rise to isometric spacetimes.
\end{abstract}

\section{Introduction}

The aim of this paper is two-fold. Firstly, we prove a geometric uniqueness result of the Characteristic Cauchy problem in General Relativity. A prerequisite for a result of this type is to have an abstract notion of the initial data completely detached from the spacetime one wishes to construct. In the standard Cauchy problem this consists of a Riemannian manifold endowed with a symmetric (0,2)-tensor satisfying the vacuum constraint equations. In the null case, not such abstract formulation was known until recently. In \cite{Mio1} we have developed a fully detached notion of initial data for the characteristic problem. The key notions to achieve this are the hypersurface data formalism, the concept of double null data and the so-called constraint tensor. However, the definition of double null data in \cite{Mio1} was given in a (partially) gauge-fixed form. The second objective of this paper is to address this problem and give a fully gauge-covariant notion of double null data. \\

The hypersurface data formalism was developed in \cite{Marc2,Marc1} to study general hypersurfaces at a purely abstract level, i.e., without reference to any ambient space. It has been employed recently in the problem of matching of spacetimes across null hypersurfaces \cite{miguel1,miguel2} and to study the characteristic problem in General Relativity \cite{Mio1}. A hypersurface data set $\mc D$ consists of an abstract manifold $\mc H$ and a set of tensors defined on $\mc H$, with which one can reconstruct the full ambient metric on the hypersurface and the transverse derivative of its tangential components whenever the data happens to be embedded. Hypersurface data has an internal gauge structure associated to the freedom in the choice of an everywhere transverse vector (so-called rigging) to the hypersurface in the embedded case. In \cite{Mio1} we have particularized the definition of hypersurface data to the null case, so that it describes an abstract null hypersurface. It turns out that that when the data is embedded, the pullback of the ambient Ricci tensor into the null hypersurface can be written completely in terms of the abstract data. This allowed us to define and study the characteristic problem for the Einstein equations in a completely abstract way. The data corresponds to two null hypersurfaces intersecting transversely in a spacetime. When the full metric is prescribed in two such null hypersurfaces, Rendall \cite{Rendall} has shown that the reduced Einstein equations are well-posed. Moreover, if the metric components are prescribed suitably, the ambient spacetime is not only a solution of the reduced equations, but also a solution of the vacuum Einstein equations. This spacetime exists in a neighbourhood of the intersection and the result is valid on any dimension and for all topologies of the intersection submanifold. Previously in \cite{friedrich1981} Friedrich considered the same problem making use of the contracted Bianchi identity, hence restricted to the four-dimensional case, and applied it both to the standard and the asymptotic characteristic problem for the Einstein equations. When the intersecting surface is a 2-sphere, Luk shows in \cite{Luk} that the spacetime can be extended to a neighbourhood of the two initial null hypersurfaces. According to \cite{rodnianski}, Luk's argument is still valid in arbitrary dimensions and with arbitrary topology of the intersection. In dimension four the existence of the solution in a full neighbourhood of the initial data hypersurfaces has been proved in \cite{cabet} for a large class of symmetric hyperbolic systems (including Einstein equations) irrespectively of the topology of the intersection. Other approaches to the characteristic problem can be found in \cite{It,CandP,Hilditch_Kroon_Zhao,Book,Yau}. In \cite{cone} this Cauchy problem is studied on the future null cone of a point. We emphasize that none of these approaches attempt to define and study the characteristic data as ``detached'' from the spacetime one wishes to construct.\\

As already mentioned, in \cite{Mio1} we approached the characteristic initial value problem in a fully abstract way by means of the hypersurface data formalism, and in particular with a new geometric object called \textit{double null data}. Our formulation detaches the initial data in the characteristic problem from the spacetime and puts the characteristic problem on the same footing as the standard Cauchy problem. A double null data set consists of two null hypersurface data $\mc D$ and $\mc{\ul D}$ with respective boundaries $\mc S$ and $\mc{\ul S}$, together with a non vanishing function at the ``corner''. In order for the double null data to be embedded in some spacetime, certain compatibility conditions at $\mc S$ must be fulfilled. Geometrically, they arise from the fact that in the embedded case the boundaries are identified. Concretely, these conditions correspond to (i) the induced metrics, (ii) the corresponding torsion one-forms and second fundamental forms, and (iii) the pullbacks of the ambient Ricci tensor be the same. As we show in \cite{Mio1}, these conditions can be written solely in terms of the abstract data. While (i) and (iii) was already written gauge-covariantly, the strategy we followed to obtain (ii) was to work in a very particular gauge so that the conditions took a simplified form. Although the existence of such gauge is always granted, defining double null data in a gauge-fixed form is not completely satisfactory from a mathematical and geometric point of view. There may be situations where writing the data in some other gauge could simplify the problem, so giving a gauge-covariant definition of double null data becomes necessary.\\

In this paper we extend the definition of double null data and make it fully gauge covariant. The construction is based on the properties of (codimension one) non-degenerate submanifolds of any given null hypersurface data. For such submanifolds we introduce the notion of normal pair, which abstractly captures the idea of a normal vector to the submanifold when embedded in an ambient space. It turns out that the (ambient) second fundamental form of the submanifold along a normal vector can be written solely in terms of the abstract data and the associated normal pair. This allows us to define, at the abstract level, the second fundamental form of a submanifold along a normal pair. In a similar way, one can define abstractly the notion of torsion one-forms associated to a basis of normal pairs. It turns out that both the second fundamental forms and the torsion one-forms are gauge-covariant, so it is natural to write the compatibility conditions for (ii) in terms of these objects. To achieve this one needs to ``glue'' abstractly the two boundaries together in a (metric) hypersurface data sense. This is accomplished by means of the notion of $\partial$-isometry, which depends on an isometry between the boundaries as well as on a non-vanishing function that geometrically corresponds to the scalar product of the null generators at the intersection whenever the data happens to be embedded. Then, the gauge-covariant definition of double null data consists of two null hypersurface data with their boundaries identified via a $\partial$-isometry and satisfying the compatibility conditions. When the double null data happens to be embedded, this notion corresponds to two transverse, null hypersurfaces, as desired.\\

The compatibility conditions are necessary for any double null data to be embeddable in some ambient spacetime. The question of whether these conditions are also sufficient arises naturally. We answer this in the affirmative by showing that given double null data there always exists a spacetime where it can be embedded. Obviously there are many spacetimes where a given double null data can be embedded and they are in general not solutions of any field equations. In that sense, the compatibility conditions are of geometric nature, and they need to be present regardless the spacetime one wants to construct is a solution of the Einstein equations (or any other field equations) or not. If in addition the abstract constraint equations hold, we proved in \cite{Mio1} that the double null data can be embedded in a spacetime solution of the $\Lambda$-vacuum Einstein equations. Here we restate this result in Theorem \ref{main} for completeness. Our strategy there was to work in the so-called harmonic gauge (which is still diffeomorphism covariant) and solve the reduced Einstein equations from the metric data. Once the spacetime is constructed we built new embedded data and show that (i) such embedded data coincides with the original one and (ii) that the spacetime is a solution of the Einstein equations. Thus, there is a clear hierarchy between the compatibility conditions and the constraint equations, in the sense that the former are purely geometric and the later depend on the field equations one wants to solve.\\

The other central question we want to address in this paper concerns the geometric uniqueness of the characteristic problem at the abstract level. In the standard Cauchy problem, it is known that two initial data sets $(\Sigma,h,K)$ and $(\Sigma',h',K')$ such that $(\Sigma,h)$ and $(\Sigma',h')$ are isometric and the isometry maps $K'$ into $K$, give rise to two isometric spacetimes $(M,g)$ and $(M',g')$ \cite{Wald}. To find a result of this type in the characteristic case we need to give a notion of isometry between two abstract initial data. To do that it is essential to take into account the gauge freedom present in the hypersurface data formalism, since two abstract hypersurface data $\mc D$ and $\mc{D}'$ related by a gauge transformation are indistinguishable from a geometric point of view. Hence, the notion of isometric double null data involves a diffeomorphism between the two abstract hypersurfaces (and their corresponding abstract tensors) as well as a gauge transformation. With this definition at hand, we prove that two double null data satisfying the constraint equations are isometric (in the abstract sense) if and only if the spacetimes they define are isometric (in the standard sense).\\

This paper is structured as follows. In Section \ref{sec_preliminares} we recall the basic definitions and results of hypersurface data from \cite{Marc2,Marc1}, such as the notion of hypersurface data, embeddedness and gauge transformation. We also summarize definitions and results from \cite{Mio1}. In Section \ref{sec_sub} we study non-degenerate codimension-one submanifolds of hypersurface data, showing that the notion of second fundamental forms and torsion one-forms of a submanifold can be defined abstractly from hypersurface data and a new object called normal pair. In Section \ref{sec_Psi} we give a fully diffeomorphism and gauge covariant definition of double null data (Def. \ref{defi_DND}), thus generalizing our previous definition in \cite{Mio1}, and we show in Theorem \ref{teo_suffi} that the compatibility conditions are necessary and sufficient for a double null data to be embeddable in some spacetime. To do that we need the explicit expression of the pullback of the constraint tensor into the boundary, which is obtained in Appendix \ref{appendix}. Finally, in Section \ref{sec_isometry}, we study the necessary conditions for two different initial data to define isometric spacetimes. This conditions lead to the definition of isometric double null data (Def. \ref{defi_isometricDND}). The paper finishes with Theorem \ref{theorem_isometry}, where we prove that two isometric double null data define isometric spacetimes. This gives a geometric uniqueness notion of the characteristic initial value problem in a fully abstract way.
	
\section*{Notation and conventions}

Let $\mc H$ be a smooth manifold. We denote by $\mc{F}(\mc H)$ the set of smooth real functions on $\mc H$, and by $\mc F^{\star}(\mc H)$ the subset of nowhere-vanishing functions. The tangent (resp. cotangent) bundle of $\mc H$ is denoted by $T\mc H$ (resp. $T^{\star}\mc H$), and the set of vector fields (resp. covector fields) is $\X(\mc H)$ (resp. $\X^{\star}(\mc H)$). The interior contraction of a tensor $T$ with a vector $X$ is $\iota_XT \d T(X,\cdots)$. Given a diffeomorphism $\phi:\mc N\to\mc M$ and a vector field $X\in\X(\mc M)$, the pullback of $X$ is given by $\phi^{\star}X \d (\phi^{-1})_{\star}X$. We employ Greek letters ($\mu,\nu,...$) for spacetime abstract indices, small Latin letters from the beginning of the alphabet ($a,b,...$) for abstract indices on $\mc H$ and capital Latin letters from the beginning of the alphabet ($A,B,...$) for abstract indices on a section of $\mc H$. Given a map $\varphi:M\to N$ and a submanifold $S$ of $M$, we denote by $\varphi|_S:S\to\varphi(S)$ the restriction of $\varphi$ to $S$, both in the domain and the image. As usual, parenthesis (resp. brackets) denote symmetrization (resp. antisymmetrization) of indices, and $\otimes_s$ is the symmetrized tensor product, i.e., $T_1\otimes_s T_2 \d \frac{1}{2}\left(T_1\otimes T_2 + T_2\otimes T_1\right)$. Our convention for the curvature tensor of a connection $\nabla$ is $R(X,Y)Z = \nabla_X\nabla_Y Z - \nabla_Y\nabla_X Z - \nabla_{[X,Y]} Z,$ and its Ricci tensor is the contraction between its contravariant index and its second covariant index. In this paper all manifolds are connected unless otherwise indicated.

\section{Preliminaries}
\label{sec_preliminares}

In this section we summarize the hypersurface data formalism. Details can be found in \cite{Marc1,Marc2} and its precursor \cite{Marc3}. This formalism has been applied in several contexts, particularly for the matching theory of spacetimes \cite{Marc1,miguel1,miguel2}. Our main interest is the characteristic case, so we shall also recall from our previous paper \cite{Mio1} the necessary definitions and results for this case.

\begin{defi}
	\label{defi_hypersurfacedata}
	Let $\mc H$ be a smooth $m$-dimensional manifold, $\bg$ a symmetric two-covariant tensor field, $\bm\ell$ a one-form and $\ell^{(2)}$ a scalar on $\mc H$. A four-tuple $\mc D^{met}=\{\mc H,\bg,\bm\ell,\ell^{(2)}\}$ defines metric hypersurface data (of dimension $m$) provided that the symmetric two-covariant tensor $\mc A|_p$ on $T_p\mc H\times\real$ defined by $$\mc A|_p\left((W,a),(Z,b)\right) \d \bg|_p (W,Z) + a\bm\ell|_p(Z)+b\bm\ell|_p(W)+ab\ell^{(2)}|_p$$ is non-degenerate at every $p\in\mc H$. A five-tuple $\mc D=\mc D^{met}\cup \{\bY\}$, where $\bY$ is a symmetric two-covariant tensor field on $\mc H$, is called hypersurface data.
\end{defi}

Since the tensor $\mc A$ is non-degenerate one can define its contravariant ``inverse'' version by $\mc{A}^{\sharp}\left(\mc A \left((V,a),\cdot\right),\cdot\right) = (V,a)$ for every $(V,a)\in\X(\mc H)\times\mc F(\mc H)$. From $\mc A^{\sharp}$ one defines the two-contravariant, symmetric tensor field $P$, the vector $n$ and the scalar function $n^{(2)}$ on $\mc H$ by means of $$\mc A^{\sharp}\left((\bm\alpha,a),(\bm\beta,b)\right) = P (\bm\alpha,\bm\beta)+a n(\bm\beta)+bn(\bm\alpha)+ab n^{(2)},$$ for all $\bm\alpha,\bm\beta\in\X^{\star}(\mc H)$ and $a,b\in\mc F(\mc H)$. Alternatively, $P$, $n$ and $n^{(2)}$ can be defined by
\begin{align}
	\bg(n,\cdot)+n^{(2)}\bm\ell&=0,\label{gamman}\\
	\bm\ell(n)+n^{(2)}\ell^{(2)}&=1,\label{ell(n)}\\
	P(\bm\ell,\cdot)+\ell^{(2)} n&=0,\label{Pell}
\end{align}
\vspace{-1cm}
\begin{align}
	P\left(\bg(X,\cdot),\cdot\right)  &= X - \bm\ell(X)n \qquad \forall X\in\X(\mc H),\label{Pgamma}\\
	\bg\left(P(\bm\alpha,\cdot),\cdot\right) & = \bm\alpha - \bm\alpha(n)\bm\ell \qquad\forall \bm\alpha\in\X^{\star}(\mc H)\label{gammaP}.
\end{align}

Despite its name, the abstract manifold $\mc H$ in Definition \ref{defi_hypersurfacedata} is not a hypersurface of any ambient space. The connection with the standard notion of hypersurface is given in the following definition.

\begin{defi}
	\label{defi_embedded}
	A metric hypersurface data $\mc D^{met}=\{\mc H,\bg,\bm\ell,\ell^{(2)}\}$ is embedded in a semi-Riemannian manifold $(\mc M,g)$ if there exists an embedding $\Phi:\mc H\hookrightarrow\mc M$ and a vector field $\xi$ along $\Phi(\mc H)$ everywhere transversal to $\Phi(\mc H)$, called rigging, such that
	\begin{equation}
		\label{embedded_equations}
		\Phi^{\star}(g)=\bg, \hspace{0.5cm} \Phi^{\star}\left(g(\xi,\cdot)\right) = \bm\ell,\hspace{0.5cm} \Phi^{\star}\left(g(\xi,\xi)\right) = \ell^{(2)}.
	\end{equation}
	The hypersurface data $\mc D=\{\mc H,\bg,\bm\ell,\ell^{(2)},\bY\}$ is embedded provided that, in addition, 
	\begin{equation}
		\label{Yembedded}
		\dfrac{1}{2}\Phi^{\star}\left(\lie_{\xi} g\right) = \bY.
	\end{equation}
\end{defi}

As usual, we define the radical of $\bg$ by $$\rad(\bg)\d \left\{X\in \X(\mc H) \text{ such that } \bg(X,\cdot)=0\right\}.$$ It can be shown that the vector space $\rad(\bg)_p$ at each $p\in\mc H$ is at most one-dimensional \cite{Marc2}. Then, from equation \eqref{gamman}, the condition $n^{(2)}=0$ is equivalent to $\rad(\gamma)=\mbox{span}\left\{ n\right\}$. When the data is embedded, the first relation in \eqref{embedded_equations} together with $\rad(\gamma)\neq \left\{ 0 \right\}$ imply that $\Phi(\mc H)$ is a null hypersurface. Motivated from this geometric picture, hypersurface data satisfying $n^{(2)}=0$ everywhere on $\mc H$ will be called \textbf{null hypersurface data} (NHD). A smooth submanifold $\mc S\subset\mc H$ is called a \textbf{section} of $\mc H$ provided that every integral curve of $n$ intersects transversely $\mc S$ exactly once.\\

Let $\mc D$ be embedded hypersurface data. Given a rigging $\xi$, any other vector of the form $\xi' = z(\xi+\zeta)$, with $z\in\mc F^{\star}(\mc H)$ and $\zeta\in\X(\mc H)$, is again a rigging vector. At the abstract level, this freedom is encoded in the following definition.

\begin{defi}
\label{defi_gauge}
Let $\mc D=\{\mc H,\bg,\bm\ell,\ell^{(2)},\bY\}$ be hypersurface data. Let $z\in\mc{F}^{\star}(\mc H)$ and ${\zeta\in\X(\mc H)}$. The gauge transformed hypersurface data with gauge parameters $(z,\zeta)$ are 
\begin{align}
	\mc{G}_{(z,\zeta)}\left(\bg \right)&\d \bg,\label{transgamma}\\
	\mc{G}_{(z,\zeta)}\left( \bm{\ell}\right)&\d z\left(\bm{\ell}+\bg(\zeta,\cdot)\right),\label{tranfell}\\
	\mc{G}_{(z,\zeta)}\left( \ell^{(2)} \right)&\d z^2\left(\ell^{(2)}+2\bm\ell(\zeta)+\bg(\zeta,\zeta)\right),\label{transell2}\\
	\mc{G}_{(z,\zeta)}\left( \bY\right)&\d z \bY + \bm\ell\otimes_s d z +\dfrac{1}{2}\lie_{z\zeta}\bg.\label{transY}
\end{align}
\end{defi}

The gauge transformation laws \eqref{transgamma}-\eqref{transell2} induce the following transformations on the contravariant data \cite{Marc1}
\begin{align}
	\mc{G}_{(z,\zeta)}\left(P \right) &= P + n^{(2)}\zeta\otimes\zeta-2\zeta\otimes_s n,\label{gaugeP}\\
	\mc{G}_{(z,\zeta)}\left( n \right)&= z^{-1}(n-n^{(2)}\zeta),\label{transn}\\
	\mc{G}_{(z,\zeta)}\left( n^{(2)} \right)&= z^{-2}n^{(2)}.
\end{align}
We will often employ a prime to denote the gauge transformed objects when the gauge group element is clear. The set of gauge transformations defines a group with composition law and inverse given by \cite{Marc1}
\begin{align}
	\mc{G}_{(z_1,\zeta_1)}\circ \mc{G}_{(z_2,\zeta_2)} &= \mc{G}_{(z_1 z_2,\zeta_2+z_2^{-1}\zeta_1)}\label{group}\\
	\mc{G}_{(z,\zeta)}^{-1}& = \mc{G}_{(z^{-1},-z\zeta)},\label{gaugelaw}
\end{align} 
and neutral element $\mc G_{(1,0)}$. Given hypersurface data we define the following two-covariant tensor fields
\begin{equation}
	\label{def_F}
	\bF\d \dfrac{1}{2}d\bm\ell
\end{equation}
and
\begin{equation}
	\label{defK}
	\bK\d n^{(2)} \bY +\dfrac{1}{2}\lie_n\bg + \bm{\ell}\otimes_s d n^{(2)}.
\end{equation} 
When the data is embedded the tensor $\bK$ corresponds \cite{Marc2} to the second fundamental form of $\Phi(\mc H)$ w.r.t. the unique normal vector $\nu$ satisfying $g(\nu,\xi)=1$. As expected from this geometric interpretation, the transformation law of $\bK$ is 
\begin{equation}
	\label{Ktrans}
	\mc{G}_{(z,\zeta)}\left(\bK \right)= z^{-1}\bK.
\end{equation} 
Given hypersurface data, a unique torsion-free connection $\ol\nabla$ exists with the following defining properties
\begin{align}
\left(\ol\nabla_X\bg\right)(Z,W) &= - \bm\ell(Z) \bK(X,W)- \bm\ell(W) \bK(X,Z),\label{olnablagamma}\\
\left(\ol\nabla_X\bm{\ell}\right)(Z)& = \bY(X,Z) + \bF(X,Z)-\ell^{(2)} \bK(X,Z).\label{olnablaell}
\end{align}

Under a gauge transformation with gauge parameters $(z,\zeta)$, the connection $\ol\nabla$ transforms as \cite{Mio1}
\begin{equation}
	\label{transfnabla}
\mc G_{(z,\zeta)}\left(\ol\nabla\right) = \ol\nabla + \zeta\otimes\bK.
\end{equation}
The combination $\bY+\bF$ will appear frequently below, so we give it a name,
\begin{equation}
	\label{defPi}
	\bPi \d \bY+\bF.
\end{equation}
The tensor $\bPi$ has the following interesting property \cite[Eq. (36)]{Mio1}
\begin{equation}
	\label{cartan}
	\bPi(n,X) - \bPi(X,n) = \left(\lie_n\bm\ell\right)(X).
\end{equation}
The transformation law of $\bPi(\cdot,n)$ is \cite[Eq. (38)]{Mio1}
\begin{equation}
	\label{transPiXn}
\bPi'(\cdot,n')  = \bPi(\cdot,n) - \bK(\cdot,\zeta) + d\log|z|.
\end{equation}

A consequence of \eqref{olnablagamma}-\eqref{olnablaell} together with \eqref{gamman}-\eqref{Pgamma} is \cite{Marc1}
\begin{equation}
	\label{olnablan}
	\ol\nabla_X n  = P\left(\bK(X,\cdot)-n^{(2)} \bPi(X,\cdot),\cdot\right)-\left(\bPi(X,n) +n^{(2)} X\left( \ell^{(2)}\right) \right)n.
\end{equation}
For null hypersurface data $(n^{(2)}=0$) the previous equation takes the simpler form 
\begin{equation}
	\label{olnablannull}
	\ol\nabla_X n  = P\left(\bK(X,\cdot),\cdot\right)-\bPi(X,n) n.
\end{equation}
The connection $\ol\nabla$ has the following geometric interpretation in the embedded case. Given $\mc D$ with embedding $\Phi$ and rigging $\xi$ in an ambient manifold $(\mc M,g)$ with Levi--Civita connection $\nabla$ and $X,Z\in\X(\mc H)$, the relation between $\ol\nabla$ and $\nabla$ is \cite{Marc1,Marc3}
\begin{equation}
	\label{nablambient}
	\nabla_X Z = \ol\nabla_X Z - \bK(X,Z)\xi,
\end{equation}
where we have identified vector fields in $\X(\mc H)$ and their image under $\Phi_{\star}$. This slight abuse of notation will be used repeatedly from now on.\\

The rest of this section is devoted to summarizing the results of characteristic hypersurface data developed in \cite{Mio1} needed in this paper. 

\begin{defi}
\label{def_CHD}
Let $\mc D=\{\mc H,\bg,\bm\ell,\ell^{(2)},\bY\}$ be hypersurface data of dimension $m$. We say that the set $\mc D$ is ``characteristic hypersurface data'' (CHD) provided that 
\begin{enumerate}
	\item $\rad(\bg)\neq\{0\}$ and $\bg$ is semi-positive definite.
	\item There exists $\ul u\in\mc{F}(\mc H)$ satisfying $\lambda\d n(\ul u)\neq 0$. Such functions are called ``foliation functions'' (FF).
	\item The leaves $\mc S_{\ul u} \d \left\{p\in\mc H: \ \ul u(p)=\ul u\right\}$ are all diffeomorphic.
\end{enumerate}
\end{defi}

Henceforth, we will assume $\mc H$ to have boundary of the form $\partial\mc H=\{\ul u=\ul u_0\}$. Given CHD, the existence of such foliation function allows us to define a tangent space decomposition at every $p\in\mc H$ of the form 
\begin{equation}
	\label{decomposition}
	T_p\mc H = T_p\mc S_{\ul u(p)}\oplus \mbox{span}\left\{ n|_p\right\},
\end{equation}
where $T_p\mc S_{\ul u(p)} = \{X\in T_p\mc H \ \text{ such that }\ X(\ul u)=0\}$. This decomposition induces another on the cotangent space given by $T_p^{\star}\mc H = T_p^{\star}\mc S_{\ul u(p)}\oplus \mbox{span}\left\{ d\ul u|_p\right\}$. Then the tangent (resp. cotangent) bundle decomposes as the direct sum $T\mc H= T\mc S \oplus \mbox{span}\left\{ n\right\}$ (resp. $T^{\star}\mc H= T^{\star}\mc S \oplus \mbox{span}\left\{ d \ul u \right\}$), where $T\mc S=\bigcup T_q\mc S_{\ul u}$, and therefore every $X\in\X(\mc H)$ can be written uniquely as $X =  X_{\para} + \sigma_X n$ with $X_{\para}\in\X(\mc S)$ and $\sigma_X\in\mc F(\mc H)$. A vector field $X$ is said to be tangent to the foliation provided $\sigma_X=0$.\\

Let $\mc D$ be CHD, $\ul u$ a foliation function and $i:\mc S_{\ul u}\hookrightarrow\mc H$ the inclusion of each leaf onto $\mc H$. For each $\mc S_{\ul u}$ we define $h\d i^{\star}\bg$, the one-form 
\begin{equation}
	\label{defellpara}
	\bm\ell_{\para}\d i^{\star}\bm\ell\in\X^{\star}(\mc S_{\ul u}),
\end{equation} 
and the vector 
\begin{equation}
	\label{defellsharp}
\ell^{\sharp}\d h^{\sharp}(\bm\ell_{\para},\cdot).
\end{equation}
It is immediate to show that $h$ is a positive definite metric. The transformation laws of $\bm\ell_{\para}$ and $\ell^{\sharp}$ follow directly from that of $\bm\ell$ (see \eqref{tranfell})
\begin{align}
\bm\ell_{\para}' &= z\left(\bm\ell_{\para} + h(\zeta_{\para},\cdot)\right),\label{transbmellpara}\\
\ell'{}^{\sharp} &= z(\ell^{\sharp} + \zeta_{\para}).\label{transellsharp}
\end{align}

%For future convenience we introduce the one-form $\bm\pi\in\X^{\star}(\mc H)$ by 
%\begin{equation}
%	\label{pi}
%	\bm\pi\d \bPi(\cdot,n) + \iota_{\ell^{\sharp}} \bK.
%\end{equation}

In \cite{Mio1} the following set of so-called ``foliation tensors'' were introduced. By construction they are intrinsic to the leaves of the foliation.
\begin{defi}
	\label{defi_foliationtensors}
Let $\mc D$ be CHD and $\ul u$ a foliation function. We define the ``foliation tensors'' $\bm\chi$, $\bm\Upsilon$, $\bt$ and $\bm\eta$ on each leaf $\mc S_{\ul u}$ by
\begin{align*}
\bm\chi\d i^{\star} \bK,\qquad \bm\Upsilon \d i^{\star} \bY,\qquad \bt \d i^{\star}\left(\bPi(\cdot,n) + \iota_{\ell^{\sharp}} \bK\right)\, \qquad \bm\eta \d -\bm{\tau} - d(\log|\lambda|).
\end{align*}
\end{defi}
%We also introduce the scalar function $\omega$ by 
%\begin{equation}
%	\omega\d \bY(n,n).
%\end{equation}	
In the embedded case, $\bm\chi$ is the second fundamental form of $\mc S_{\ul u}$ along the normal vector $\nu$. The tensors $\bm\Upsilon$ and $\bm\eta$ do not admit such a neat geometric interpretation. However, in an appropriate gauge (namely such that the rigging is null and normal to $\mc S_{\ul u}$), $\bm\Upsilon$ coincides with the null second fundamental form along $\xi$ and $\bm\eta$ is the torsion one-form of $\mc S_{\ul u}$ w.r.t the null basis $\{\nu,\xi\}$. Finally we recall that $\bY(n,n)$ measures the deviation of $\nu$ from being geodesic (this follows exclusively from equations \eqref{olnablaell}, \eqref{nablambient} and therefore it does not depend on the gauge or on the existence of a foliation on $\mc H$). We refer the reader to \cite[Sect. 5]{Mio1} for the details.\\

As shown in \cite{Mio1}, given embedded null hypersurface data in a spacetime $(\mc M,g)$, all tangential components of the ambient Ricci tensor on the hypersurface can be written in terms of the abstract data. This allowed us to define the abstract constraint tensor $\bm{\mc R}_{ab}$ by 
\begin{align}
	\bm{\mc R}_{ab}&\d \left(\bg_{af}\ol{R}^f{}_{cbd} + 2\ol\nabla_{[d}\left(\bK_{b]c}\bm\ell_a\right)+2\ell^{(2)} \bK_{c[b}\bK_{d]a}\right)P^{cd}\label{Rab}\\
	&\quad\, - \left(\bm\ell_d\ol{R}^d{}_{bac} + 2\ell^{(2)}\ol\nabla_{[c}\bK_{a]b} + \bK_{b[a}\ol\nabla_{c]}\ell^{(2)} + \bm\ell_d\ol{R}^d{}_{abc} + 2\ell^{(2)}\ol\nabla_{[c}\bK_{b]a} + \bK_{a[b}\ol\nabla_{c]}\ell^{(2)}\right)n^c,\nonumber
\end{align}
where $\ol{R}^a{}_{bcd}$ is the curvature tensor of $\ol\nabla$. As shown in \cite{Mio1}, this tensor satisfies 
\begin{equation}
	\label{Ricciembedded}
	\bm{\mc R} = \Phi^{\star}\ric,
\end{equation} 
where $\ric$ is the Ricci tensor of $(\mc M,g)$. This property suggests that $\bm{\mc R}$ is gauge invariant and, indeed, in Theorem 4.6 in \cite{Mio1} we have established
\begin{equation}
	\label{gaugeRab}
	\mc{G}_{(z,\zeta)}\left(\bm{\mc R}\right) = \bm{\mc R} \qquad \forall (z,\zeta)\in\mc{F}^{\star}(\mc H)\times\X(\mc H).
\end{equation}

By the geometric interpretation of this tensor, the condition $\bm{\mc R}=0$ can be thought as the vacuum constraint equations on a null hypersurface $\mc H$. These equations are fully diffeomorphism and gauge covariant and do not assume the presence of any foliation, thus generalizing the so-called null structure equations. Recall that the hypersurface data formalism allows us to write these constraint equations even though there is no metric on $\mc H$, thanks to the existence of the abstract connection $\ol\nabla$.\\

In Sect. 6 of \cite{Mio1} we introduced a fully covariant gauge which translates the harmonic condition on the ambient coordinates into an abstract condition on the data. Given a set of independent functions $\{x^{\ul a}\}$ on $\mc H$ and a (local) basis $\{e_c\}$ of $\X(\mc H)$, we define $B_c{}^{\ul a}\d e_c(x^{\ul a})$ and $B^c{}_{\ul a}$ as its inverse. Since, for each value of $\ul a$, $B^c{}_{\ul a}$ is a vector so it is $V^c \d B^c{}_{\ul a} \square_P x^{\ul a}$, where $\square_P f \d P^{ab}\ol\nabla_a\ol\nabla_b f$ for every scalar $f$. This vector allows us to define the so-called harmonic gauge, which one can show it exists and is unique up to the choice of the gauge parameters at a given ``initial'' section. We refer the reader to Theorem 6.2 of \cite{Mio1} for the proof.

\begin{teo}
	\label{teo_HG}
Let $\mc D$ be null hypersurface data admitting a section $\mc S\subset \mc H$, $\{x^{\ul a}\}$ a set of $m$ functionally independent functions, $V^c\d B^c{}_{\ul a}\square_P x^{\ul a}$ and $\tr_P\bK\d P^{ab}\bK_{ab}$. Then there exists a class of gauges satisfying the following conditions on $\mc H$,
	\begin{align}
		\tr_P \bK -2\bY(n,n)&=0,\label{combGamma}\\
		2P\left(\ol\nabla_n\bm\ell,\cdot\right)+n\left(\ell^{(2)}\right)n&=V.\label{V^c2}
	\end{align}
The set of transformations keeping the previous conditions invariant is parametrized by $(z_0,\zeta_0)\in\mc F^{\star}(\mc S)\times\X(\mc H)$. Any gauge satisfying \eqref{combGamma} and \eqref{V^c2} will be called ``harmonic gauge'' (HG).
\end{teo}

The name ``harmonic'' is motivated by the following property. Given embedded CHD in a spacetime $(\mc M,g)$ with rigging $\xi$ and embedding $\Phi$, we can extend the functions $x^{\ul a}$ off $\Phi(\mc H)$ by $\xi(x^{\ul a})=0$. Moreover, we can introduce another function $u$ on $\mc M$ such that $u|_{\mc H}=0$ and $\xi(u)=1$ on $\Phi(\mc H)$. As shown in \cite{Mio1}, the data is written in the harmonic gauge if and only if the functions $\{u,x^{\ul a}\}$ satisfy $\square_g x^{\ul a} = \square_g u=0$ on $\Phi(\mc H)$.\\

One can exploit the remaining gauge freedom in the class of gauges of Theorem \ref{teo_HG} to fix the values of $\ell^{(2)}$ and $\bm\ell_{\para}$ to zero in any section $\mc S$ (see \cite[Cor. 4.4]{Mio1}).

\begin{lema}
	\label{HG2}
Let $\mc D$ be null hypersurface data admitting a section $\mc S\subset \mc H$, $\{x^{\ul a}\}$ a set of $m$ functionally independent functions and $V^c\d B^c{}_{\ul a}\square_P x^{\ul a}$. Then there exists a class of gauges satisfying \eqref{combGamma}-\eqref{V^c2} on $\mc H$ as well as $\ell^{(2)}=0$, $\bm\ell_{\para}=0$ on $\mc S$. The set of transformations keeping the previous conditions invariant is parametrized by $z_0\in\mc F^{\star}(\mc S)$.
%\begin{proof}
%Let $z\in\mc F^{\star}(\mc S)$ and $\zeta=i_{\star}\zeta_{\para}+\sigma n \in\X(\mc H)$, $\zeta_{\para}\in\X(\mc S)$, $\sigma\in\mc F(\mc S)$. Directly from the transformation laws \eqref{transell2} and \eqref{transbmellpara},
%\begin{align}
%z^2\left(\ell^{(2)}+2\bm\ell_{\para}(\zeta_{\para})+2\sigma + h\left(\zeta_{\para},\zeta_{\para}\right)\right)&=0,\label{eq1}\\
%z\left(\bm\ell_{\para}+h(\zeta_{\para},\cdot)\right)&=0.\label{eq2}
%\end{align}
%Equation \eqref{eq2} admits a unique solution for $\zeta_{\para}$, which introduced into \eqref{eq1} fixes $\sigma$. 
%\end{proof}
\end{lema}

Let $\{x^{\ul a}\}=\{\ul u, x^A\}$ be functions on $\mc H$ satisfying (i) $n(\ul u)\neq 0$, (ii) $n(x^A)=0$ and (iii) $\{x^A\}$ is a coordinate system on $\mc S$. When the data is written in the gauge of Lemma \ref{HG2} w.r.t $\{x^{\ul a}\}$, conditions \eqref{V^c2} can be rewritten as equations involving the $\bY$ tensor (see \cite[Prop. 7.13]{Mio1}).

\begin{prop}
	\label{propHG3}
	Let $\mc D$ be null hypersurface data admitting a section $\mc S$ written in the gauge of Lemma \ref{HG2} w.r.t some functions $x^{\ul a}=\{\ul u, x^A\}$ satisfying conditions (i), (ii) and (iii) above. Then the following equations hold at $\mc S$
	\begin{align}
\tr_h \bm\Upsilon \d h^{AB}\bm\Upsilon_{AB} &= n\left(\ell^{(2)}\right),\label{eqHG1}\\
2\left(\lie_n\bm\ell + \bPi(\cdot,n)\right)(\grad_h x^A)  & = \square_h x^A.\label{eqHG2}
	\end{align}
\end{prop}

\section{Non-degenerate submanifolds}

\label{sec_sub}

In this section we study embedded non-degenerate submanifolds in the context of hypersurface data. For the purposes of this paper we restrict ourselves to the null case but without fixing a priori the topology of $\mc H$. Let $\mc D = \{\mc H,\bg,\bm\ell,\ell^{(2)},\bY\}$ be null hypersurface data of dimension $m$ and $\Sigma$ an $(m-1)$-dimensional manifold. We say that $i:\Sigma\hookrightarrow\mc H$ is a non-degenerate submanifold (of codimension one) of $\mc D$ provided that $h\d i^{\star}\bg$ is non-degenerate. The idea is to identify, in terms of hypersurface data, a suitable notion of the torsion and second fundamental form of $\Sigma$. We start by assuming that $\mc D$ is embedded null hypersurface data with rigging $\xi$ and embedding $\Phi$ in a spacetime $(\mc M,g)$ with Levi-Civita connection $\nabla$. Given a vector field $t$ along $\Phi\left(i\left(\Sigma\right)\right)$ everywhere normal to $\Sigma$, there exist a function $C\in\mc{F}(\Sigma)$ and a vector field $T\in\X(\mc H)$ such that 
\begin{equation}
	\label{vectort}
	t = C \xi + \Phi_{\star} T.
\end{equation} 
By \eqref{embedded_equations}, the condition that $t$ is normal to $\Sigma$ is equivalent to $$ C\bm\ell(X) + \bg(T,X)=0 \qquad \forall X\in\X(\Sigma).$$ Given $X,Y\in\X(\Sigma)$, the (ambient) second fundamental form $\II^t$ of $\Sigma$ along the normal vector $t$ can be defined as $$\II^t (X,Y) = \dfrac{1}{2}\left(\lie_t g\right)(X,Y).$$ Substituting the expression \eqref{vectort} and employing the simple formula
\begin{equation}
\label{formula}
\lie_{fV} g = f\lie_V g + 2d f\otimes_s g(V,\cdot)
\end{equation} 
valid for any function $f$ and vector field $V$, gives $$\II^t(X,Y) = \dfrac{1}{2}\left(\lie_{\Phi_{\star}T} g\right)(X,Y) + \dfrac{1}{2} C\left(\lie_{\xi} g \right)(X,Y) + \left(d C\otimes_s g(\xi,\cdot)\right)(X,Y).$$ Inserting the expressions of $\bY$ and $\bm\ell$ in the context of embedded hypersurface data (see Def. \ref{defi_embedded}) and using $\Phi^{\star}\left(\lie_{\Phi_{\star} T} g \right) = \lie_T\left( \Phi^{\star} g\right) =\lie_T\bg$, one finally gets 
\begin{equation}
\label{Kambient}
\II^t (X,Y) = \dfrac{1}{2}\left(\lie_T\bg\right)(X,Y) + C \bY(X,Y) + \dfrac{1}{2}\left(X(C)\bm\ell(Y) + Y(C)\bm\ell(X)\right),
\end{equation}
which is an expression only depending on (abstract) hypersurface data.\\

 In addition to the second fundamental form, the geometry of submanifolds of (ambient) codimension two also consist of the torsion one-form. Given a basis of normal vectors $\{t_i = C_i \xi +\Phi_{\star} T_i\}$, $i=1,2$, the (ambient) torsion of $\Sigma$ w.r.t. this basis is the set of one-forms $T_{ij}$ given by 
\begin{equation}
	\label{daleth}
T_{ij}(X) = g(t_i,\nabla_X t_j) \qquad \forall X\in\X(\Sigma).
\end{equation} 
As in the case of the second fundamental form, this object can be also rewritten in terms of hypersurface data. Indeed, from \eqref{vectort} and \eqref{nablambient}
\begin{equation}
\label{aux4}
\begin{aligned}
	\nabla_X t_j & = \nabla_X\left(C_j\xi +\Phi_{\star}T_j\right) \\
	&=\left( X(C_j) - \bK(X,T_j)\right) \xi + C_j \nabla_X\xi + \ol\nabla_X T_j.
	\end{aligned}
\end{equation}
Then, using \eqref{embedded_equations} 
\begin{equation}
	\label{aux5}
	g(\xi,\nabla_X t_j) = \left(X(C_j)-\bK(X,T_j)\right)\ell^{(2)} + \dfrac{1}{2} C_j X(\ell^{(2)}) + \bm\ell(\ol\nabla_X T_j),
\end{equation} 
and $$g(\Phi_{\star}T_i,\nabla_X t_j) = \left(X(C_j)-\bK(X,T_j)\right)\bm\ell(T_i) + C_j g (\Phi_{\star}T_i,\nabla_X\xi) + \bg(T_i,\ol\nabla_X T_j).$$ From equations \eqref{nablambient}, \eqref{olnablaell} and the definition of $\bm\ell$ in the context of embedded data (Def. \ref{defi_embedded}), 
\begin{align*}
g(\Phi_{\star} T_i,\nabla_X\xi) &= X\left( g(\Phi_{\star}T_i,\xi)\right) - g\left(\nabla_X (\Phi_{\star} T_i),\xi\right) \\
&= X\left( g(\Phi_{\star}T_i,\xi)\right) -  g\left(\ol\nabla_X T_i,\xi\right) + \bK(X,T_i) g(\xi,\xi) \\
&=X\left(\bm\ell(T_i)\right) - \bm\ell\left(\ol\nabla_X T_i\right)+ \ell^{(2)} \bK(X,T_i) \\
&=\left(\ol\nabla_X\bm\ell\right)(T_i) + \ell^{(2)} \bK(X,T_i)\\
& = \bPi(X,T_i).
\end{align*}
Then,
\begin{equation}
	\label{aux6}
g(\Phi_{\star}T_i,\nabla_X t_j) = \left(X(C_j)-\bK(X,T_j)\right)\bm\ell(T_i) + C_j \bm\Pi(X,T_i) + \bg(T_i,\ol\nabla_X T_j).
\end{equation}
Introducing \eqref{aux4} into \eqref{daleth} and employing \eqref{aux5} and \eqref{aux6},
\begin{equation}
	\label{daleth2}
\begin{aligned}
T_{ij}(X) & = g(C_i\xi + \Phi_{\star}T_i,\nabla_X t_j)\\
&=\left(\bm\ell(T_i) + C_i\ell^{(2)}\right) \left(X(C_j)-\bK(X,T_j)\right) + C_i\bm\ell\left(\ol\nabla_X T_j\right) \\
&+ \bg(T_i,\ol\nabla_X T_j) + \dfrac{1}{2} C_i C_j X\left(\ell^{(2)}\right) + C_j \bPi(X,T_i).
\end{aligned}
\end{equation}

The computation above suggests introducing a fully abstract notion of normal vector to the submanifold. This notion is called ``normal pair'' (NP) and it is defined as follows.
\begin{defi}
\label{def_NP}
	Let $\mc D=\{\mc H,\bg,\bm\ell,\ell^{(2)},\bY\}$ be hypersurface data and $i:\Sigma\hookrightarrow\mc H$ a non-degenerate submanifold. A normal pair $\mf t\d \{T,C\}$ consists of a vector field $T\in\X(\mc H)$ along $i(\Sigma)$ and a function $C\in\mc F(\Sigma)$ satisfying 
	\begin{equation}
		\label{eq_NP}
		C\bm\ell(X) + \bg(T,X)=0 \qquad \forall X\in\X(\Sigma).
	\end{equation}
\end{defi}
Given that $\Sigma$ is a codimension one submanifold of $\mc H$, $\Sigma$ admits exactly two linearly independent NPs. Indeed, equation \eqref{eq_NP} can be written in terms of the tensor $\mc A$ (see Def. \ref{defi_hypersurfacedata}) as $$\mc A\left((T,C),(X,0)\right)=0$$ for every $X\in \X(\Sigma)$ along $i(\Sigma)$. Since $\mc A|_p$ is non-degenerate at every $p\in\Sigma$, the $(m+1)$-dimensional vector space $T_p\mc H\times \real$ can be decomposed as $$T_p\mc H\times\real = T_p\Sigma \oplus \left(T_p\Sigma\right)^{\bot}.$$ Since $T_p\Sigma$ has dimension $(m-1)$ it follows that $\left(T_p\Sigma\right)^{\bot}$ is two-dimensional and hence $\Sigma$ admits exactly two linearly independent NPs. Motivated by \eqref{Kambient}, given a normal pair $\mf t=\{T,C\}$, the second fundamental form of $\Sigma$ along $\mf t$ is the tensor field on $\Sigma$ defined by 
\begin{equation}
\label{defi_Kt}
	\bm{\mc{K}}^{\mf t} (X,Y) \d \dfrac{1}{2}\left(\lie_T\bg\right)(X,Y) + C \bY(X,Y) + \dfrac{1}{2}\left(X(C)\bm\ell(Y) + Y(C)\bm\ell(X)\right),
\end{equation}
for $X,Y\in\X(\Sigma)$. In the embedded case, given a normal pair $\mf t = \{C,T\}$ we can associate to it the (ambient) vector field $t[\mf t]$ defined by $$t[\mf t] \d C\xi + \Phi_{\star} T.$$ By construction, $t[\mf t]$ is normal to $\Phi(i(\Sigma))$, since $g(t[\mf t],X)=C\bm\ell(X) + \bg(T,X)= 0$ for every $X\in\X(\Sigma)$. Comparing the expression of the (ambient) second fundamental form $\II^{t[\mf t]}$ of $\Phi\left(i(\Sigma)\right)$ in \eqref{Kambient} with the abstract $\bm{\mc{K}}^{\mf t}$ in definition \eqref{defi_Kt}, the following result follows.
\begin{prop}
Let $\mc D=\{\mc H,\bg,\bm\ell,\ell^{(2)},\bY\}$ be embedded hypersurface data on a spacetime $(\mc M,g)$ with embedding $\Phi$ and rigging $\xi$, and let $\Sigma$ be a non-degenerate submanifold of $\mc H$. Let $\mf t=\{T,C\}$ be a NP of $\Sigma$ and let $t=C\xi+\Phi_{\star}T$ be its associated (ambient) vector field. Then, $$\bm{\mc{K}}^{\mf t}(X,Y) = \II^{t[\mf t]}(X,Y)$$ for every $X,Y\in\X(\Sigma)$.
\end{prop}

As expected from the spacetime interpretation, $\bm{\mc{K}}^{\mf t}$ has the following properties.

\begin{lema}
	\label{lema_escala}
	Let $\mf t = \{T,C\}$ be a normal pair, $\bm{\mc{K}}^{\mf t}$ the second fundamental form as in \eqref{defi_Kt} and $f$ a scalar function. Then, $$\bm{\mc{K}}^{f\mf t} = f\bm{\mc{K}}^{\mf t}.$$
	\begin{proof}
Directly from the definition \eqref{defi_Kt} and the formula \eqref{formula},
\begin{align*}
\bm{\mc{K}}^{f\mf t}(X,Y) & = \dfrac{1}{2}\left(\lie_{fT}\bg\right)(X,Y) + fC \bY(X,Y) + \left(d \left(fC\right)\otimes_s\bm\ell\right)(X,Y)\\
	& = f\bm{\mc{K}}^{\mf t}(X,Y) + \left(d f\otimes_s \bg(T,\cdot)\right)(X,Y) + \left(Cd f \otimes_s \bm\ell\right)(X,Y)\\
	&=f\bm{\mc{K}}^{\mf t}(X,Y) +\left(d f\otimes_s \left(\bg(T,\cdot) + C\bm\ell\right)\right)(X,Y).
\end{align*}
Since $\mf t$ is a normal pair, the term in parenthesis in the last line vanishes when contracted with tangent vectors, and thus result follows.
	\end{proof}
\end{lema}

\begin{lema}
	\label{lema_Kbasis}
Let $\{\mf t_i\}$ be a basis of normal pairs and $\hat{\mf t}_i = \Omega_i^j\mf t_j$ a change of basis, with $\Omega^j_i\in\mc F(\Sigma)$. Then, $$\bm{\mc{K}}^{\hat{\mf t}_i} = \Omega_i^j \bm{\mc{K}}^{\mf t_i}.$$
\begin{proof}
By the previous lemma it suffices to show that $\bm{\mc K}^{\mf t_1 + \mf t_2}=\bm{\mc K}^{\mf t_1} + \bm{\mc K}^{\mf t_2}$ for every pair of NPs $\mf t_1=(T_1,C_1)$ and $\mf t_2=(T_2,C_2)$. Directly from the definition of $\bm{\mc K}$ in \eqref{defi_Kt},
\begin{align*}
\bm{\mc K}^{\mf t_1 + \mf t_2}(X,Y) & = \dfrac{1}{2}\left(\lie_{T_1+T_2}\bg\right)(X,Y) + (C_1+C_2) \bY(X,Y)\\
&\quad + \dfrac{1}{2}\left(X(C_1+C_2)\bm\ell(Y) + Y(C_1+C_2)\bm\ell(X)\right)\\
&= \bm{\mc K}^{\mf t_1}(X,Y) + \bm{\mc K}^{\mf t_2}(X,Y).
\end{align*}
Finally, $\bm{\mc K}^{\mf t_1 + \mf t_2}= \bm{\mc K}^{\mf t_1} + \bm{\mc K}^{\mf t_2}$ together with $\bm{\mc{K}}^{f\mf t} = f\bm{\mc{K}}^{\mf t}$ proves $\bm{\mc{K}}^{\hat{\mf t}_i} = \Omega_i^j \bm{\mc{K}}^{\mf t_i}$.
\end{proof}
\end{lema}

The spacetime picture also motivates the following definition.

\begin{defi}
	\label{defi_NPGauge}
	Let $\mc D=\{\mc H,\bg,\bm\ell,\ell^{(2)},\bY\}$ be hypersurface data and $i:\Sigma\hookrightarrow\mc H$ a non-degenerate submanifold. Let $(z,\zeta)$ be a gauge group element. The gauge transformation of a normal pair $\mf t = \{T,C\}$ of $\Sigma$ is defined as $$\mc G_{(z,\zeta)}(\mf t)\d\{T' = T-C\zeta,C' = z^{-1}C\}.$$
\end{defi} 

Note that $\mf t'=\mc G_{(z,\zeta)}(\mf t)$ is still a normal pair, since $$C' \bm\ell' (X) + \bg(T',X) = C\bm\ell(X) + C\bg(\zeta,X) + \bg(T,X) - C\bg(\zeta,X) = 0.$$ Moreover, Def. \ref{defi_NPGauge} is a realization of the gauge group. Indeed, let $(z_1,\zeta_1),(z_2,\zeta_2)$ be gauge parameters. On the one hand,
$$\mc G_{(z_1,\zeta_1)}\mc G_{(z_2,\zeta_2)} \{T,C\} = \mc G_{(z_1,\zeta_1)} \left\{ T-C\zeta_2,z_2^{-1}C\right\} = \left\{ T-(z_2^{-1}\zeta_1 + \zeta_2)C,z_1^{-1}z_2^{-1} C\right\},$$ and on the other, using \eqref{group}, $$\mc G_{(z_1,\zeta_1)}\mc G_{(z_2,\zeta_2)} \{T,C\} = \mc G_{(z_1z_2,\zeta_2+z_2^{-1}\zeta_1)} \left\{T,C\right\} = \left\{T-(z_2^{-1}\zeta_1 + \zeta_2)C,z_1^{-1}z_2^{-1} C\right\}.$$ As expected from the spacetime picture, the second fundamental form along a normal pair is gauge invariant.

\begin{lema}
	\label{lema_gaugeKt}
Let $\mc D$ be hypersurface data, $\Sigma$ a non-degenerate submanifold and $\mf t$ a normal pair of $\Sigma$. Then $\bm{\mc{K}}^{\mf t}$ is gauge invariant.
\begin{proof}
Let $\mf t=\{T,C\}$ and $\mf t'=\{T'=T-C\zeta,C'=z^{-1}C\}$. Using the transformation laws \eqref{tranfell} and \eqref{transY} (we drop the argument $(X,Y)$ in order not to overload the notation)
\begin{align*}
\bm{\mc{K}}^{\mf t'}  = &\,\, \dfrac{1}{2}\lie_{T'}\bg + C' \bY' + d C' \otimes_s\bm\ell'\\
 = &\,\, \dfrac{1}{2}\lie_T\bg - \dfrac{1}{2}C\lie_{\zeta}\bg - d C\otimes_s\bg(\zeta,\cdot) + C\bY + z^{-1} C \bm\ell\otimes_s d z + \dfrac{1}{2} C\lie_{\zeta}\bg\\
& + z^{-1} C d z\otimes_s \bg(\zeta,\cdot) + d C\otimes_s\left(\bm\ell + \bg(\zeta,\cdot)\right) + zC d z^{-1}\otimes_s \left(\bm\ell + \bg(\zeta,\cdot)\right)\\
=&\,\, \dfrac{1}{2}\lie_T\bg  + C\bY  + d C\otimes_s\bm\ell\\
=&\,\, \bm{\mc{K}}^{\mf t},
\end{align*}
where in the second line we used the formula \eqref{formula}.
\end{proof}
\end{lema}

Let $i:\Sigma\hookrightarrow\mc H$ be a non-degenerate submanifold. Following the notation in \eqref{defellpara}-\eqref{defellsharp} in the context of CHD, we define the one-form $\bm\ell_{\para}\in\X^{\star}(\Sigma)$ and the vector $\ell^{\sharp}\in\X(\Sigma)$ by means of
\begin{equation}
	\label{objects}
\bm\ell_{\para}\d i^{\star}\bm\ell, \qquad \ell^{\sharp}\d h^{\sharp}(\bm\ell_{\para},\cdot).
\end{equation}

In the next proposition we identify a particular relevant normal pair.

\begin{prop}
Let $\mc D=\{\mc H,\bg,\bm\ell,\ell^{(2)},Y\}$ be null hypersurface data and $i:\Sigma\hookrightarrow\mc H$ a non-degenerate submanifold. Define the vector $\theta$ along $i(\Sigma)$ by 
\begin{equation}
	\label{theta}
	\theta\d -\dfrac{1}{2}\left(\ell^{(2)}-h\left(\ell^{\sharp},\ell^{\sharp}\right)\right) n - i_{\star}\ell^{\sharp}.
\end{equation}
Then, $\mf t_{\theta}\d (\theta,1)$ is a normal pair of $\Sigma$. Furthermore, $\mf t_{\theta}$ is $\mc A$-null, i.e., $\mc A(\mf t_{\theta},\mf t_{\theta})=0$.
\begin{proof}
For any $X\in\X(\Sigma)$, $$\mc A\left((i_{\star}X,0),(\theta,1)\right) = \bm\ell(i_{\star}X) + \bg(i_{\star}X,\theta) = \bm\ell_{\para}(X) - h(X,\ell^{\sharp}) = 0,$$ where we used $\bg(n,\cdot)=0$. Hence, $\mf t_{\theta}$ is a NP. Moreover, $$\mc{A}\left((\theta,1),(\theta,1)\right) = \bg(\theta,\theta) + 2\bm\ell(\theta) + \ell^{(2)} = h(\ell^{\sharp},\ell^{\sharp})  -\ell^{(2)}+h\left(\ell^{\sharp},\ell^{\sharp}\right) - 2\bm\ell_{\para}(\ell^{\sharp}) +\ell^{(2)} = 0,$$ where now we used $\bg(n,\cdot)=0$ and $\bm\ell(n)=1$.
\end{proof}
\end{prop}

\begin{rmk}
	\label{anal}
Observe that the normal pair $\mf t_n\d (n,0)$ is linearly independent of $\mf t_{\theta}$ and also satisfies $\mc A(\mf t_n,\mf t_n)=0$. In fact, since the space of normal pairs is two-dimensional and $$\mc A(\mf t_n,\mf t_{\theta})=\bg(n,\theta) + \bm\ell(n) =1,$$ it follows that any normal pair which is also $\mc A$-null lies either in $\mbox{span}\{\mf t_n\}$ or $\mbox{span}\{\mf t_{\theta}\}$.
\end{rmk}

 From equation \eqref{daleth2} the following abstract definition of the torsion one-forms is motivated.

\begin{defi}
	\label{def_daleth}
Let $\mc D=\left\{\mc H,\bg,\bm\ell,\ell^{(2)},\bY\right\}$ be null hypersurface data and $\left\{\mf t_i = (T_i,C_i)\right\}$ with $i=1,2$, a basis of normal pairs. The torsion one-forms of $\Sigma$ w.r.t this basis are the tensors on $\Sigma$ defined by 
\begin{align*}
	\daleth_{ij}(X) & \d \left(\bm\ell(T_i) + C_i\ell^{(2)}\right) \left(X(C_j)-\bK(X,T_j)\right) + C_i\bm\ell\left(\ol\nabla_X T_j\right) \\
	& + \bg(T_i,\ol\nabla_X T_j) + \dfrac{1}{2} C_i C_j X\left(\ell^{(2)}\right) + C_j \bPi(X,T_i)
\end{align*}
for every $X\in\X(\Sigma)$. We will also employ the notation $\daleth(\mf t_i,\mf t_j)$ when $\daleth_{ij}$ may cause confusion.
\end{defi} 

Let $\{\mf t_i,\mf t_j\}$ be a basis of normal pairs and let $t_i[\mf t_i]$, $t_j[\mf t_j]$ be their associated (ambient) vector fields. Comparing the expression of the (ambient) torsion one-forms $T_{ij}$ of $\Phi\left(i(\Sigma)\right)$ in \eqref{daleth} with the abstract $\daleth_{ij}$ in Definition \ref{def_daleth}, the following result follows.
\begin{prop}
Let $\mc D=\{\mc H,\bg,\bm\ell,\ell^{(2)},\bY\}$ be embedded hypersurface data on a spacetime $(\mc M,g)$ with embedding $\Phi$ and rigging $\xi$, and let $\Sigma$ be a non-degenerate submanifold of $\mc H$ with a basis of normal pairs $\{\mf t_i\}$. Then, $$\daleth_{ij}(X) = T_{ij}(X)$$ for every $X\in\X(\Sigma)$, where $T_{ij}$ are the ambient torsion one-forms w.r.t the basis $\{t_i[\mf t_i]\}$.
\end{prop}

Since $\Sigma$ is a codimension-one submanifold of $\mc H$ one has in principle four different torsion one-forms $\daleth_{ij}$. However, not all them are independent.

\begin{prop}
	\label{prop_daleths}
Let $\mc D=\left\{\mc H,\bg,\bm\ell,\ell^{(2)},\bY\right\}$ be null hypersurface data, $\left\{(T_i,C_i)\right\}$ a basis of normal pairs and 
\begin{equation}
	\label{Mij}
	M_{ij}\d \mc A\left((T_i,C_i),(T_j,C_j)\right) =\bg(T_i,T_j)+ C_i \bm\ell(T_j) + C_j \bm\ell(T_i) + C_i C_j \ell^{(2)}.
\end{equation} 
Then, $$\daleth_{ij} + \daleth_{ji} = d M_{ij}.$$
\begin{proof}
Directly from the definition of $\daleth$,
\begin{equation}
	\label{aux1}
	\begin{aligned}
		\daleth_{ij}(X) + \daleth_{ji}(X) & = X\left(C_i C_j \ell^{(2)}\right) + 2\left( X(C_{(i}) - \bK(X, T_{(i}) \right)\bm\ell(T_{j)}) - 2\ell^{(2)} C_{(i} \bK(X,T_{j)})\\
		&\quad + 2C_{(i}\bm\ell\left(\ol\nabla_X T_{j)}\right) + 2 \bg\left(T_{(i},\ol\nabla_X T_{j)}\right) + 2 C_{(i}\bPi(X,T_{j)}).
	\end{aligned}
\end{equation}
Using equation \eqref{olnablagamma}, 
\begin{equation}
	\label{aux2}
	2 \bg\left(T_{(i},\ol\nabla_X T_{j)}\right) = X\left(\bg(T_i,T_j)\right) - \left(\ol\nabla_X\bg\right)(T_i,T_j) = X\left(\bg(T_i,T_j)\right) +2 \bK(X,T_{(i})\bm\ell(T_{j)}).
\end{equation} 
From \eqref{olnablaell}, 
\begin{equation}
	\label{aux3}
	\begin{aligned}
2 C_{(i}\bm\ell\left(\ol\nabla_X T_{j)}\right) & = 2X\left(C_{(i}\bm\ell(T_{j)})\right) - 2X(C_{(i})\bm\ell(T_{j)}) - 2C_{(i}\left(\ol\nabla_X\bm\ell\right)(T_{j)})\\
& = 	2X\left(C_{(i}\bm\ell(T_{j)})\right) - 2X(C_{(i})\bm\ell(T_{j)}) - 2C_{(i} \bPi(X,T_{j)}) + 2 \ell^{(2)} C_{(i} \bK(X, T_{j)}).
\end{aligned}
\end{equation}
Inserting \eqref{aux2} and \eqref{aux3} into \eqref{aux1} yields the result.
\end{proof}
\end{prop}
In fact, since $\Sigma$ is a codimension one submanifold of $\mc H$, there is only one independent torsion one-form on $\Sigma$, that can be taken to be $\daleth_{12}$. In the embedded case, given two normal pairs $\mf t_i$ and $\mf t_j$, the function $M_{ij}=\mc A\left(\mf t_i,\mf t_j\right)$ as defined in \eqref{Mij} corresponds to the scalar product $g(t_i,t_j)$, where $t_i$ and $t_j$ are the ambient vector fields associated to $\mf t_i$ and $\mf t_j$, respectively. As expected from this geometric interpretation, the functions $M_{ij}$ are gauge invariant, as we show next at the abstract level.

\begin{lema}
	\label{Mijgauge}
Let $\{\mf t_i\}$ be a basis of normal pairs of a non-degenerate submanifold $\Sigma$ and $M_{ij} \d \mc A\left(\mf t_i,\mf t_j\right)$. Then, the functions $M_{ij}$ are gauge invariant.
\begin{proof}
Let $(z,\zeta)$ be gauge parameters. Directly from \eqref{Mij} and Def. \ref{defi_NPGauge},
\begin{align*}
	M'_{ij} & = \mc{A}'\left( (T_i-C_i\zeta,z^{-1}C_i),(T_j-C_j\zeta,z^{-1}C_j)\right)\\
	& = \bg\left(T_i-C_i\zeta,T_j-C_j\zeta\right) + z^{-1}C_i\bm\ell'\left(T_j-C_j\zeta\right) + z^{-1}C_j\bm\ell'\left(T_i-C_i\zeta\right) + z^{-2}C_i C_j\ell^{(2)}{}'\\
	&= \bg\left(T_i,T_j\right) + C_i\left(z^{-1}\bm\ell'(T_j) - \bg(\zeta,T_j)\right)+ C_j\left(z^{-1}\bm\ell'(T_i) - \bg(\zeta,T_i)\right) \\
	&\quad + C_iC_j\left(z^{-2}\ell^{(2)}{}' + \bg(\zeta,\zeta) - 2z^{-1}\bm\ell'(\zeta)\right)\\
	&=\bg(T_i,T_j)+C_i\bm\ell(T_j) + C_j\bm\ell(T_i)+C_i C_j \ell^{(2)}\\
	&=M_{ij},
\end{align*}
where in the fourth line we used \eqref{transgamma}-\eqref{transell2}.
\end{proof}
\end{lema}

 As expected from the geometric interpretation of the ambient torsion one-forms as the connection coefficients of the normal bundle connection, the transformation of the abstract torsion one-forms under a change of basis of normal pairs is as follows.

\begin{prop}
	\label{basis_NPs}
	Let $\{\mf t_i\}$ be a basis of normal pairs of $\Sigma$ and $\hat{\mf t}_j = \Omega_j^i\mf t_i$ with $\Omega_j^i\in \mc F(\Sigma)$ a change of basis. Then,
	\begin{equation}
		\label{transdaleth0}
		\wh\daleth_{ij} = \Omega_i^k\Omega_j^l \daleth_{kl} + \Omega_i^k M_{kl}\ d \Omega_j^l.
	\end{equation}
\begin{proof}
Writing $\mf t_i = (T_i,C_i)$ and $\hat{\mf t}_i  = (\wh T_i,\wh C_i)$, the change of basis gives $$\wh C_i = \Omega_i^k C_k,\qquad \wh T_i = \Omega_i^k T_k.$$ Inserting them in the expression of $\daleth_{ij}$ and noting that all the terms are multilinear except for $X(\wh C_j)$ and $\ol\nabla_X \wh T_j$, which become $$X(\wh C_j) = \Omega^l_j X(C_l) + C_l X(\Omega^l_j),\qquad \ol\nabla_X \wh T_j = \Omega_j^l \ol\nabla_X T_l + X(\Omega^l_j) T_l,$$ one immediately gets $$\wh\daleth_{ij} (X) = \Omega_i^k\Omega_j^l \daleth_{kl} + \Omega_i^k \left(C_k C_l \ell^{(2)} + C_k \bm\ell(T_l) + C_l \bm\ell(T_k) +\bg(T_k,T_l)\right)  X\left(\Omega_j^l\right),$$ which is \eqref{transdaleth0} after recalling the definition of $M_{kl}$ in \eqref{Mij}.
\end{proof}
\end{prop}

\begin{cor}
	\label{transdaleth}
	Let $\mf t_1$ and $\mf t_2$ two linearly independent normal pairs and $f$ a scalar function. Under a change of basis $\{\mf t_1,\mf t_2\}\mapsto \{\mf t_1'=f^{-1}\mf t_1,\mf t_2'=f \mf t_2\}$, the torsion one-form $\daleth(\mf t_1,\mf t_2)$ transforms as $$\wh\daleth_{12}  = \daleth_{12} + M_{12} d\log|f|.$$
%	\begin{proof}
%		From the definition of $\daleth$,
%		\begin{align*}
%			\daleth_{12}' (X) - \daleth_{12}(X) & = \left(\bm\ell(T_1)+C_1\ell^{(2)}\right)C_2 f^{-1} X(f)  + C_1 \bm\ell(T_2) f^{-1} X(f) + \bg(T_1,T_2) f^{-1} X(f) \\
%			& = \left(C_1 C_2 \ell^{(2)} + C_1\bm\ell(T_2) + C_2\bm\ell(T_1) + \gamma(T_1,T_2)\right) X(\log|f|) .
%		\end{align*}
%		Taking into account \eqref{Mij}, the result follows.
%	\end{proof}
\end{cor}

In the next proposition we show that the abstract torsion one-forms are gauge invariant.

\begin{prop}
	\label{torsion_gauge}
	Let $\mc D$ be hypersurface data, $\Sigma$ a non-degenerate submanifold and $\{\mf t_i=(T_i, C_i)\}$ a basis of normal pairs. Then, the torsion one-forms $\daleth_{ij}$ are gauge invariant.
	\begin{proof}
		First we show that if $\daleth_{ij}$ are gauge invariant in a particular basis of normal pairs, they are also invariant in any basis. Let $\{\mf t_{i}\}$ (with $i=1,2$) be a basis of normal pairs and let $\hat{\mf t}_{j} = \Omega_{j}^{i}\mf t_{i}$ with $\Omega_j^i \in \mc F(\Sigma)$ be a change of basis. From the transformation law in Def. \ref{defi_NPGauge}, $$\wh C_j = \Omega_j^i C_i, \quad \wh T_j = \Omega_j^i T_i \quad\Longrightarrow\quad z^{-1} \wh C_j = \Omega_j^i z^{-1} C_i,\quad \wh T_j-\wh C_j\zeta = \Omega_j^i \left(T_i - C_i\zeta\right),$$ which means that the gauge transformed basis $\{\mf{t}'_i\}$ and $\{\hat{\mf t}_i'\}$ are related by $\hat{\mf t}_j' = \Omega^i_j \mf{t}_i'$. Thus, the functions $\Omega_j^i$ are gauge invariant. Moreover, from Lemma \ref{Mijgauge} the functions $M_{ij}$ are gauge invariant too. Then, from Proposition \ref{basis_NPs}, if $\daleth_{ij}$ is gauge invariant, so it is $\wh\daleth_{ij}$. Hence, it suffices to show the statement in the basis $\{\mf t_n,\mf t_{\theta}\}$, where the only non-zero torsion one-forms are given by $$\daleth_{n\theta}(X)=-\daleth_{\theta n}(X) = -\bK(X,\theta)+\bm\Pi(X,n).$$ From Def. \ref{defi_NPGauge}, given gauge parameters $(z,\zeta)$ the transformed basis of normal pairs is $\mc G\left(\mf t_n\right) = (n,0)$ and $\mc G\left(\mf t_{\theta}\right) = \left(\theta-\zeta,z^{-1}\right)$. Hence, applying Def. \ref{def_daleth} to the new gauge,
		\begin{align*}
			\mc G\left(\daleth_{n\theta}\right)(X) & = z\left(X(z^{-1}) - \bK'(X,\theta-\zeta)\right) + z^{-1}\bPi'(X,n)\\
			& = -X\left(\log|z|\right) - \bK(X,\theta) + \bK(X,\zeta) + \bPi'(X,n')\\
			&=-X\left(\log|z|\right) - \bK(X,\theta) + \bK(X,\zeta) + \bPi(X,n) + X\left(\log|z|\right)-\bK(X,\zeta)\\
			&=\daleth_{n\theta}(X),
		\end{align*}
		where in the third line we used the transformation law of $\bPi(\cdot,n)$ in \eqref{transPiXn}.
	\end{proof}
\end{prop}

\begin{rmk}
	\label{remark}
	In the context of CHD the leaves $\{ i_{\ul u}\hspace{-0.1cm}:S_{\ul u}\hookrightarrow\mc H\}$ are non-degenerate submanifolds of $\mc H$. Then, it is natural to connect the geometry of the foliation in a CHD with the tools developed in this Section. Let $\mc D$ be CHD, $\ul u$ a foliation function and $\mc S_{\ul u}$ any leaf. By Remark \ref{anal}, $\mf t_n = (n,0)$ and $\mf t_{\theta}= (\theta,1)$ constitute a basis of $\mc A$-null, normal pairs of $\mc S_{\ul u}$. Using \eqref{defi_Kt} and Definition \ref{defi_foliationtensors}, we have $\bm{\mc{K}}^{\mf t_n} = \bm\chi$ and $\bm{\mc{K}}^{\mf t_{\theta}} = \bm\Upsilon + \bm\Theta$, where 
	\begin{equation}
		\label{Theta}
		\bm\Theta \d \dfrac{1}{2}i_{\ul u}^{\star}\left(\lie_{\theta}\bg\right).
	\end{equation} 
	From the definition of the torsion one-form (Def. \ref{def_daleth}) one also has $$\daleth_{n\theta} (X) = -\bK(X,\theta)+\bPi(X,n) = \bK(X,\ell^{\sharp}) + \bPi(X,n) = \bm\tau(X).$$ Thus, the tensors $\bm\chi$ and $\bm\Upsilon + \bm\Theta$ are the second fundamental forms w.r.t the normal pairs $\mf{t}_n = (n,0)$ and $\mf{t_{\theta}}=(\theta,1)$, respectively, whereas the tensor $\bt$ is the torsion one-form w.r.t the basis $\{\mf t_n,\mf t_{\theta}\}$.
\end{rmk}

\section{Gauge-covariant compatibility conditions}

\label{sec_Psi}

In Section 7 of \cite{Mio1} we studied the necessary conditions that two CHD must fulfil in order to be simultaneously embedded in the same spacetime with common spacelike boundary. However, these compatibility conditions were obtained in a particular gauge where strong restricting conditions were imposed at the respective boundaries. In this Section we generalize the compatibility conditions and write them employing the tools developed in the previous section, which allows us to define the notion of double null data in a fully gauge-covariant way. We start introducing the key notion that will allow us to do that.

\begin{defi}
	\label{def_isometry}
Let $\mc D$ and $\mc{\ul D}$ be two null hypersurface data with boundaries $\mc S\d\partial\mc H$ and $\ul{\mc S}\d\partial\mc{\ul H}$ and ${\phi:\mc S\to\mc{\ul S}}$ a diffeomorphism between them. An invertible linear map $$\Psi_p: T_p\mc H\oplus\real\to T_{\phi(p)}\mc{\ul H}\oplus\real$$ is called a $\partial$-isometry at $p\in\mc S$ provided that\footnote{Given $V,W\in T_p\mc H\oplus\real$, we define $\big(\Psi^{\star}\mc{\ul A}_{\phi(p)}\big)(V,W)\d \mc{\ul A}_{\phi(p)}\left(\Psi(V),\Psi(W)\right)$.}
	\begin{enumerate}
		\item $\Psi_p|_{T_p\mc S}\left((X,0)\right) = \left(\phi_{\star}|_p(X),0\right)$ for all $X\in\X(\mc S)$,
		\item $\Psi_p^{\star}\mc{\ul A}_{\phi(p)} = \mc A_p$.
	\end{enumerate}
If conditions (1) and (2) hold at every $p\in\mc S$ we say that $\mc D$ and $\mc{\ul D}$ are $\partial$-isometric.
\end{defi}
Since $\mc{S}$ and $\mc{\ul S}$ are non-degenerate codimension one submanifolds of $\mc H$ and $\mc{\ul H}$, we can talk about normal pairs on $\mc{S}$ and $\mc{\ul S}$. Given a NP $\mf t=\{T,C\}$ of $\mc S$, the image of $\mf t$ under $\Psi$ is still a normal pair on $\mc{\ul S}$, since $\phi$ is a diffeomorphism and $$0 = \mc A\left(\mf t,(X,0)\right)=\left(\Psi^{\star}\ul{\mc A}\right)\left(\mf t,(X,0)\right) = \mc{\ul A}\left(\Psi(\mf t),(\phi_{\star} X,0)\right) \qquad \forall X\in\X(\mc H).$$ Moreover, if a normal pair $\mf t$ of $\mc{S}$ is $\mc A$-null, the normal pair $\Psi(\mf t)$ of $\mc{\ul S}$ is $\ul{\mc A}$-null, since $$ \mc{\ul A} \left(\Psi(\mf t),\Psi(\mf t)\right) = \mc A\left(\mf t,\mf t\right)=0.$$ In the following proposition we show that given an isometry $\phi:\mc{S}\to\mc{\ul S}$ and a non-vanishing function $\mu\in\mc F^{\star}(\mc{\ul S})$, a natural $\partial$-isometry can be constructed.

\begin{prop}
	\label{prop_uniqueness}
Let $\mc D$ and $\mc{\ul D}$ be two NHD and $\phi:(\mc{S},h)\to(\mc{\ul S},\ul h)$ an isometry. For each $\mu\in\mc F^{\star}(\mc{\ul S})$, there exists a unique $\partial$-isometry $\Psi: T\mc H\oplus\mc F(\mc S)\to T\mc{\ul H}\oplus\mc F(\mc{\ul S})$ determined by the condition $\ul{\mc A}\left((\ul n,0), \Psi(n,0)\right) = \mu$. 
	\begin{proof}
The existence and uniqueness proof will be constructive, i.e., we will impose conditions (1) and (2) in Def. \ref{def_isometry} and show that there is a unique candidate. We will then check that this candidate is indeed a $\partial$-isometry. Let $X\in\X(\mc{S})$. We define $\Psi\left((X,0)\right)\d \left(\phi_{\star} X,0\right)$. In order to define the image of $(n,0)$ and $(\theta,1)$ under $\Psi$ first observe that since $(n,0)$ and $(\theta,1)$ are $\mc A$-null, $\Psi\left((n,0)\right)$ and $\Psi\left((\theta,1)\right)$ are $\mc{\ul A}$-null too, and since $\ul{\mc A}\left((\ul n,0),(\ul\theta,1)\right)=1$ (see Remark \ref{anal}) we necessarily have $\Psi\left((n,0)\right)\in \mbox{span}\left\{(\ul\theta,1)\right\}$ and $\Psi\left((\theta,1)\right) \in\mbox{span}\left\{(\ul n,0)\right\}$. Imposing the condition $\mc{\ul A}\left((\ul n,0),\Psi(n,0)\right)=\mu$ the only option is
\begin{equation}
	\label{B(n)}
	\Psi\left((n,0)\right) = \mu(\ul\theta,1).
\end{equation} 
To complete the determination of $\Psi$ we only need to find $\Psi\left((\theta,1)\right)$, which we already know is of the form $\Psi\left((\theta,1)\right) = \alpha (\ul n,0)$. The proportionality function $\alpha$ is obtained from $\mc{A}\left((n,0),(\theta,1)\right)=1$ and its underlined version after imposing item (2) of Def. \ref{def_isometry} as follows $$1=\mc A\left((\theta,1),(n,0)\right)=\mc{\ul A}\left(\Psi\left((\theta,1)\right),\Psi((n,0))\right) = \alpha \mu \mc{\ul A}\left((\ul n,0),(\ul\theta,1)\right)=\alpha\mu.$$ Since $\mu\neq 0$ by hypothesis we conclude
\begin{equation}
	\label{Psitheta}
	\Psi\left((\theta,1)\right) = \mu^{-1} (\ul n,0).
\end{equation}

In matrix notation, the expression of $\Psi$ in the decomposition $T\mc{S}\oplus\mbox{span}\left\{(n,0),(\theta,1)\right\}$ (and the corresponding one in the image) is\footnote{When a map is written in matrix notation we follow the convention that the entries of the $i$-th column are the coefficients of the image of the $i$-th vector in the basis.}
\begin{equation}
	\label{matrix}
	\Psi = \left(\begin{matrix}
		\left(\phi_{\star}\right) &\bm 0 & \bm 0 \\
		\bm{0} & 0 & \mu^{-1}\\
		\bm{0} & \mu & 0
	\end{matrix}\right).
\end{equation}

It is now straightforward to check that this candidate to be a $\partial$-isometry is indeed a {$\partial$-isometry}. Indeed, conditions (1) and (2) of Def. \ref{def_isometry} hold by construction and because $\phi$ is an isometry, and by expression \eqref{matrix} and the fact that $\phi$ is a diffeomorphism, we conclude that $\Psi$ is invertible.
	\end{proof}
\end{prop}

In the proof of Prop. \ref{prop_uniqueness} we found the expression of $\Psi$ w.r.t decompositions $T_p\mc{S}\oplus\mbox{span}\left\{ (n,0),(\theta,1)\right\}$ and $T_{\phi(p)}\mc{\ul S}\oplus\mbox{span}\left\{ (\ul n,0),(\ul\theta,1)\right\}$, namely
\begin{equation}
	\label{matrixPsi0}
	\Psi = \left(\begin{matrix}
		\left(\phi_{\star}\right) &\bm 0 & \bm 0 \\
		\bm{0} & 0 & \mu^{-1}\\
		\bm{0} & \mu & 0
	\end{matrix}\right).
\end{equation}
In some situations it may be interesting to have an expression for $\Psi$ in a more general basis. 

\begin{prop}
Let $\mc D$ and $\mc{\ul D}$ be NHD, $\phi:\mc{S}\to\mc{\ul S}$ an isometry and $\Psi$ the unique $\partial$-isometry satisfying $\ul{\mc A}\left((\ul n,0), \Psi(n,0)\right) = \mu\neq 0$. Let $v$, $\ul v$ be transverse vectors to $\mc{S}$ and $\mc{\ul S}$, respectively. Then the linear map $\Psi$ w.r.t decompositions $T\mc{S}\oplus\mbox{span}\left\{(v,0),(0,1)\right\}$ and $T\mc{S}\oplus\mbox{span}\left\{(\ul v,0),(0,1)\right\}$ is given by 
\begin{equation}
\label{matrixPsi}
\Psi = \left(\begin{matrix}
\left(\phi_{\star}\right) & \phi_{\star}v_{\para} -\sigma\mu\left(\ubar\alpha\ubar\sigma^{-1}\ul v_{\para} + \ul\ell^{\sharp}\right) & \phi_{\star}\ell^{\sharp} +\mu\alpha\ul\ell^{\sharp} + \ul\sigma^{-1}(\mu\alpha\ul\alpha-\mu^{-1})\ul v_{\para} \\
\bm{0} & \mu\sigma\ubar\alpha\ubar\sigma^{-1} & \ul\sigma^{-1}(\mu^{-1}-\mu\alpha\ul\alpha)\\
\bm{0} & \mu\sigma & - \mu \alpha
		\end{matrix}\right),
	\end{equation}
	where $\sigma\in\mc F^{\star}(\mc{S})$, $\ul\sigma\in\mc F^{\star}(\mc{\ul S})$, $v_{\para}\in\X(\mc{S})$ and $\ul{v}_{\para}\in\X(\mc{\ul S})$ are univocally defined by the decompositions $v = \sigma n + v_{\para}$ and $\ul v = \ubar\sigma \ubar n + \ul v_{\para}$, and where we introduce the functions $\alpha=-\frac{1}{2}\left(\ell^{(2)}-h(\ell^{\sharp},\ell^{\sharp})\right)$ and $\ul\alpha=-\frac{1}{2}\big(\ul\ell^{(2)}-\ul h(\ul\ell^{\sharp},\ul\ell^{\sharp})\big)$ to simplify the notation.
	\begin{proof}
		We only need to compute the second and third columns. Since $\Psi\left((v,0)\right) = \sigma \Psi\left((n,0)\right) + \Psi\left((v_{\para},0)\right)$, employing \eqref{B(n)} together with \eqref{theta}, namely $\ul\theta = \ubar\alpha\ubar n - \ul\ell^{\sharp}$ (we omit the $i_{\star}$ in order not to overload the notation), $$\Psi\left((v,0)\right) = \sigma\mu(\ul\theta,1) + (\phi_{\star}v_{\para},0) = \left(\sigma\mu\ubar\alpha\ubar\sigma^{-1}(\ul v-\ul v_{\para}) - \sigma\mu\ul\ell^{\sharp} +\phi_{\star}v_{\para},\sigma\mu\right),$$ and hence the second column of \eqref{matrixPsi} follows. The third column requires computing $\Psi\left((0,1)\right)$. Decomposing $(0,1)=(\theta,1)-(\alpha n,0)+(\ell^{\sharp},0)$ and using \eqref{matrixPsi0} yields
\begin{align*}
\Psi(0,1) &= \left(\mu^{-1}\ul n -\mu\alpha \ul \theta + \phi_{\star}\ell^{\sharp},-\mu\alpha\right)\\
&=\left(\mu^{-1}\ul\sigma^{-1}(\ul v-\ul v_{\para}) - \mu\alpha\ubar\alpha\ubar\sigma^{-1}(\ul v-\ul v_{\para}) + \mu\alpha\ul\ell^{\sharp}+\phi_{\star}\ell^{\sharp},-\mu\alpha\right),
\end{align*}
so the third column of \eqref{matrixPsi} is established.
	\end{proof}
\end{prop}

Next we study how the map $\Psi$ changes under gauge transformations on $\mc D$ and $\ul{\mc D}$.

\begin{prop}
	\label{proptransPsi}
Let $\mc D$ and $\mc{\ul D}$ be null hypersurface data and $\Psi$ be a $\partial$-isometry. Under gauge transformations on $\mc D$ and $\mc{\ul D}$ with parameters $(z,\zeta)$ and $(\ul z,\ul\zeta)$, respectively, $\Psi$ transforms as 
	\begin{equation}
		\label{transfB}
\Psi' = \ul G^{-1} \circ \Psi\circ G,
	\end{equation}
where $G$ is the invertible linear map
\begin{align}
	G: T\mc H\oplus\mc F(\mc{S}) &\longrightarrow T\mc H\oplus\mc F(\mc{S})\nonumber\\
	(V,a) &\mapsto G\left((V,a)\right) \d \left(V+az\zeta,az\right)\label{G}
\end{align}
and $\ul G: T\mc{\ul H}\oplus\mc F(\mc{\ul S}) \longrightarrow T\mc{\ul H}\oplus\mc F(\mc{\ul S})$ is defined identically but with all the quantities carrying an underline. As a consequence, the transformation law of the function $\mu\d \mc {\ul A}\left((\ul n,0),\Psi(n,0)\right)$ is
\begin{equation}
	\label{transmu}
	\mu' = z^{-1}\ul z^{-1}\mu,
\end{equation}
where in order not to overload the notation we denote with the same symbol a function $f\in \mc F(\partial \mc H)$ and $f\circ \phi^{-1} \in \mc F(\partial \mc{\ul H})$. This slight abuse of notation will be used repeatedly from now on when no confusion arises.
\begin{proof}
Let $\mc D$ be hypersurface data. From Def. \ref{defi_gauge} one can write the transformation law for the tensor $\mc A$ in a compact way as
\begin{equation}
	\label{transA2}
	\mc A' \left((V,a),(W,b)\right) = \mc A\left(G(V,a),G(W,b)\right)
\end{equation}
for all $(V,a),(W,b)\in T\mc H\oplus \mc F(\mc{S})$. To show this we use matrix notation in which vectors are represented as columns and covectors as rows. The map \eqref{G} gets rewritten in matrix form as $$ \left(\begin{matrix}
	V+az\zeta \\
	az
\end{matrix}\right)=\left(\begin{matrix}
	\mathbb{1} & z\zeta\\
	0 & z
	\end{matrix}\right)\left(\begin{matrix}
	V \\
	a
	\end{matrix}\right).$$ Define, therefore 
\begin{equation*}
	(G) = \left(\begin{matrix}
	\mathbb{1} & z\zeta\\
	0 & z
		\end{matrix}\right),\qquad (\mc A) =  \left(\begin{matrix}
		\bg & \bm\ell^T\\
		\bm\ell & \ell^{(2)}
	\end{matrix}\right).
	\end{equation*}
Then \eqref{transA2} can be written in matrix form as
\begin{equation}
	\label{transfA}
	(\mc{A}') = (G)^T (\mc A)  (G).
\end{equation}
To prove this equality (and hence \eqref{transA2}) we compute
\begin{align*}
(G)^T (\mc A) (G) &= \left(\begin{matrix}
	\mathbb{1} & 0\\
	z\zeta^T & z
\end{matrix}\right)\left(\begin{matrix}
\bg & \bm\ell^T\\
\bm\ell & \ell^{(2)}
\end{matrix}\right)\left(\begin{matrix}
\mathbb{1} & z\zeta\\
0 & z
\end{matrix}\right)\\
& = \left(\begin{matrix}
\bg & \bm\ell^T\\
z\left(\bg(\zeta,\cdot)+\bm\ell\right) & z\left(\bm\ell(\zeta)+\ell^{(2)}\right)
\end{matrix}\right)\left(\begin{matrix}
\mathbb{1} & z\zeta\\
0 & z
\end{matrix}\right)\\
& = \left(\begin{matrix}
\bg & z\left(\bg(\zeta,\cdot)+\bm\ell\right)^T\\
z\left(\bg(\zeta,\cdot)+\bm\ell\right) & z\left(\bg(\zeta,\zeta)+2\bm\ell(\zeta)+\ell^{(2)}\right)
\end{matrix}\right),
\end{align*}
which is precisely the matrix form of $\mc A'$ after taken into account the transformation laws \eqref{transgamma}-\eqref{transell2}. Item (2) of Def. \ref{def_isometry} can be also written in matrix form as $(\Psi)^T(\mc{\ul A}) (\Psi) = (\mc{A})$. To compute the gauge behaviour of $\Psi$ we impose $(\Psi')^T (\ul{\mc A}')(\Psi') = (\mc A')$ and use equation \eqref{transfA}, 
\begin{align*}
(\Psi')^T (\mc{\ul A}') (\Psi') = (\mc{A}') &\Longleftrightarrow (\Psi')^T(\ul{G})^T (\ul{\mc A}) (\ul G) (\Psi') = (G)^T(\mc{A})(G)\\
& \Longleftrightarrow \left((G)^T\right)^{-1}(\Psi')^T(\ul G)^T (\ul{\mc A})(\ul G)(\Psi') (G)^{-1}=(\mc A),
\end{align*} 
from where we conclude that $\Psi=\ul G\circ\Psi' \circ G^{-1}$ and hence $\Psi' = \ul G^{-1}\circ \Psi\circ  G$, which is \eqref{transfB}. The transformation of $\mu$ follows from those of $\Psi$ and $\mc A$. Indeed, by \eqref{transA2}, $$\mu'\d \mc{\ul A}'\left(\Psi'(n',0),(\ul n',0)\right) = \ul{\mc A} \left(\left(\ul G\circ\Psi'\right) (n',0),\ul G(\ul n',0)\right),$$ and using \eqref{transfB} together with $G\left((n',0)\right) = z^{-1}(n,0)$ (and its underlined version),
\begin{align*}
\mu' &=\ul{\mc A} \left(\left(\Psi \circ G\right) (n',0),\ul G(\ul n',0)\right)\\
&=\ul{\mc A} \left(\Psi \left( G(n',0)\right),\ul G(\ul n',0)\right)\\
&=z^{-1}\ul z^{-1} \ul{\mc A} \left(\Psi \left((n,0)\right),(\ul n,0)\right)\\
&= z^{-1}\ul z^{-1}\mu.
\end{align*}
\vskip -8mm
\end{proof}
\end{prop}

\begin{rmk}
	\label{remark1}
The transformation law of a normal pair in Def. \ref{defi_NPGauge} can be rewritten in terms of the map $G$ defined in \eqref{G} by $\mf t_i' = G^{-1}\left(\mf t_i\right)$.
\end{rmk}

Next we want to define two null and transverse hypersurfaces from an abstract point of view, i.e., without seeing them as embedded in any ambient spacetime. In \cite{Mio1} the notion of double embedded CHD was introduced in order to study how two different CHD fit together in the same spacetime when we identify their boundaries. However, using the notion of normal pair and the map $\Psi$, it is no longer necessary to introduce such object. Indeed, let $\mc D$ and $\ul{\mc D}$ be embedded CHD with respective embeddings $\Phi$ and $\ul\Phi$ in a spacetime $(\mc M,g)$, and suppose $\Phi(\mc{S})=\ul\Phi(\mc{\ul S})\eqqcolon S$. Let $\phi:\mc{S}\to\mc{\ul S}$ be the induced diffeomorphism and let $\Psi$ be a $\partial$-isometry satisfying $\ul{\mc A}\left(\Psi(n,0),(\ul n,0)\right)=\mu\in\mc F(\mc{\ul S})$. By Defs. \ref{defi_hypersurfacedata} and \ref{defi_embedded}, the object $\ul{\mc A}\left(\Psi(n,0),(\ul n,0)\right)$ can be thought as the scalar product $g(\nu,\ul \nu)$ at $S$, where $\nu=\Phi_{\star}n$ and $\ul\nu=\ul\Phi_{\star}\ul n$. Thus, if one wants to define abstractly two null and transverse hypersurfaces with the same orientation for $\nu$ and $\ul\nu$, the function $\ul{\mc A}\left(\Psi(n,0),(\ul n,0)\right)$ must be everywhere negative when $n$ and $\ul n$ point into the interior of $\mc H$ and $\mc{\ul H}$, respectively. Then by Prop. \ref{prop_uniqueness} the condition $\ul{\mc A}\left(\Psi(n,0),(\ul n,0)\right)=\mu\neq 0$ fixes uniquely the map $\Psi$. Moreover, if we want $\Phi(\mc{S})$ and $\ul\Phi(\mc{\ul S})$ to correspond to the same (codimension two) surface in the ambient spacetime, there are several necessary conditions that they need to fulfil. Firstly, their induced metrics have to agree. Secondly, their second fundamental forms and torsion one-forms also need to agree. And finally, the pullback of the ambient Ricci tensor into the two surfaces must agree too. The ``zeroth order'' condition, namely that the induced metrics coincide, can be simply written as $\phi^{\star}\ul h = h$. In order to write the ``first order'' conditions, i.e. the ones of the second fundamental forms and torsion one-forms, we can employ the tools developed before. As seen in Section \ref{sec_sub}, these conditions can be expressed abstractly thanks to the notion of normal pairs. Indeed, identifying the ambient vectors associated to a basis of normal pairs $\{\mf t_i\}$ with the ambient vectors associated to the normal pairs $\left\{\Psi(\mf t_i)\right\}$, the second fundamental forms $\bm{\mc K}^{\mf t_i}$ and the torsions $\daleth(\mf t_i,\mf t_j)$ of $\mc{S}$ must agree with those of $\mc{\ul S}$, namely $\bm{\mc{\ul K}}^{\Psi(\mf t_i)}$ and $\daleth(\Psi(\mf t_i),\Psi(\mf t_j))$. Finally, the ``second order'' condition, i.e. the one involving the ambient Ricci tensor, can be expressed abstractly thanks to the abstract tensor $\bm{\mc R}$ defined in \eqref{Rab}. Indeed, since $\Phi^{\star}\ric = \bm{\mc R}$ and $\ul\Phi^{\star}\ric = \bm{\ul{\mc R}}$ (see \eqref{Ricciembedded}), the pullback of the tensors $\bm{\mc R}$ and $\bm{\ul{\mc R}}$ on $\mc{S}$ and $\mc{\ul S}$ must also agree. This whole discussion motivates the following abstract and fully gauge-covariant definition of double null data.

\begin{defi}
	\label{defi_DND}
Let $\mc H$ and $\mc{\ul H}$ be two manifolds with boundaries $i:\mc{S}\hookrightarrow\mc H$ and ${\ul i:\mc{\ul S}\hookrightarrow\mc{\ul H}}$, let $\phi:\mc{S}\to\mc{\ul S}$ be an isometry and $\mu\in\mc{F}(\mc{\ul S})$ everywhere negative. Let $\mc D=\{\mc H,\bg,\bm\ell,\ell^{(2)},\bY\}$ and $\mc{\ul D}=\{\mc{\ul H},\ul\bg,\bm{\ul\ell},\ul\ell^{(2)},\ul \bY\}$ be CHD and restrict $n$ and $\ul n$ to point towards the interior of $\mc H$ and $\mc{\ul H}$, respectively. Let $\Psi$ be the unique $\partial$-isometry such that $\ul{\mc A}\left(\Psi(n,0),(\ul n,0)\right)=\mu$, and $\{\mf t_i\}$ a basis of normal pairs of $\mc{S}$. Then we say that the triple $\{\mc D,\mc{\ul D},\mu\}$ is double null data (DND) provided that the following conditions hold at $\mc{S}$
	\begin{align}
		\phi^{\star}\ul h & = h,\label{conditionh}\\
	\Psi^{\star}\left(\ul{\bm{\mc{K}}}^{\Psi(\mf t_i)}\right) &= \bm{\mc{K}}^{\mf t_i},\label{condition1}\\
	\Psi^{\star}\left(\ul\daleth(\Psi(\mf t_i),\Psi(\mf t_j))\right) &=\daleth(\mf t_i,\mf t_j),\label{condition3}\\
	\phi^{\star}\left(\ul i^{\star}\bm{\mc{\ul R}}\right) &= i^{\star}\bm{\mc R}.\label{conditionR}
\end{align}
\end{defi}

Although \eqref{conditionh} is redundant since $\phi:\mc{S}\to\mc{\ul S}$ is already assumed to be an isometry, we write down it again for completeness. In the next remark we show that Def. \ref{defi_DND} is well-defined, i.e., that the compatibility conditions are independent of the basis of NPs and also gauge invariant.

\begin{rmk}
	\label{remark_DNDgaugeinvariant}
Conditions $\phi^{\star}\ul h=h$ and $\phi^{\star}\left(\ul i^{\star}\bm{\mc{\ul R}}\right) = i^{\star}\bm{\mc R}$ are automatically gauge invariant by virtue of the transformations laws \eqref{transgamma} and \eqref{gaugeRab}. Then it suffices to show the gauge invariance of \eqref{condition1}-\eqref{condition3}. We start with \eqref{condition1}. Let $(z,\zeta)$ and $(\ul z,\ul \zeta)$ be gauge parameters and denote by $\mf t'_i$ the gauge transformed normal pair of $\mf t_i$. Firstly, the RHS of \eqref{condition1} is simply $\bm{\mc K}^{\mf t'_i}$ as a direct consequence of the gauge invariance of the second fundamental form in Lemma \ref{lema_gaugeKt}. We need to show that the RHS can also be written with all objects gauge transformed. Let $G$, $\ul G$ be defined as in Prop. \ref{proptransPsi}. By Lemma \ref{lema_gaugeKt} the LHS of \eqref{condition1} can be written as $$\Psi^{\star}\left(\ul{\bm{\mc{K}}}^{\Psi(\mf t_i)}\right)=\Psi^{\star}\left(\bm{\mc{\ul K}}^{\ul G^{-1} \left(\Psi(\mf t_i)\right)}\right),$$ since the second fundamental form $\bm{\mc{\ul K}}$ along a normal pair is gauge invariant and $\ul G^{-1}\left(\Psi(\mf t_i)\right)$ is the gauge transformation of the normal pair $\Psi(\mf t_i)$ with gauge parameters $(\ul z,\ul\zeta)$ (see Remark \ref{remark1}). Then, using $\ul G^{-1}\circ \Psi = \Psi'\circ G^{-1}$ (by Prop. \ref{proptransPsi}), $$\Psi^{\star}\left(\bm{\mc{\ul K}}^{\ul G^{-1}\left( \Psi(\mf t_i)\right)}\right) = \Psi{}^{\star}\left(\bm{\mc{\ul K}}^{\Psi'\left(G^{-1}(\mf t_i)\right)}\right)=\Psi{}^{\star}\left(\bm{\mc{\ul K}}^{\left(\Psi'(\mf t'_i)\right)}\right),$$ where in the last equality we used again $\mf t_i' = G^{-1}\mf t_i$. Noting that $\Psi\left((X,0)\right)= \Psi'\left((X,0)\right)$ for every $X\in\X(\mc{\ul S})$, the LHS of equation \eqref{condition1} finally gets rewritten as $$\Psi'{}^{\star}\left(\bm{\mc{\ul K}}^{\left(\Psi'(\mf t'_i)\right)}\right),$$ which is exactly the LHS of \eqref{condition1} with all quantities gauge transformed. This establishes the gauge covariance of conditions \eqref{condition1}. Moreover, by Lemma \ref{lema_Kbasis}, \eqref{condition1} are also independent of the basis of NPs.\\

The gauge invariance of conditions \eqref{condition3} can be shown in a similar way as in the case of the second fundamental form. Firstly, the RHS of \eqref{condition3} is simply $\daleth(\mf t_i',\mf t_j')$, by the gauge invariance of $\daleth$ in Prop. \ref{torsion_gauge}. To see that the LHS of \eqref{condition3} can be written with all quantities gauge transformed, we use again $\ul G^{-1}\circ \Psi = \Psi'\circ G^{-1}$, $\mf t_i' = G^{-1}\mf t_i$ and $\Psi\left((X,0)\right)= \Psi'\left((X,0)\right)$, so that $$\Psi^{\star}\left(\ul\daleth(\Psi(\mf t_i),\Psi(\mf t_j))\right)=\Psi{}^{\star}\left(\ul\daleth(\ul G^{-1}\left(\Psi(\mf t_i)\right),\ul G^{-1}\left(\Psi(\mf t_j)\right))\right)=\Psi'{}^{\star}\left(\ul\daleth(\Psi'(\mf t'_i),\Psi'(\mf t'_j))\right),$$ where in the first equality we invoked again the gauge invariance of $\daleth$ (Prop. \ref{torsion_gauge}). Thus, the gauge covariance of \eqref{condition3} is also established. Finally, by Prop. \ref{basis_NPs} under a change of basis of NPs $\hat{\mf t}_i = \Omega_i^k\mf t_k$, the RHS of \eqref{condition3} transforms as $$\daleth(\hat{\mf t}_i,\hat{\mf t}_j) = \Omega_i^k\Omega_j^l\daleth(\mf t_k,\mf t_l) + \Omega_i^k M_{kl} d\Omega^l_j,$$ whereas the transformation of the LHS is
\begin{align*}
	\Psi^{\star}\left(\ul\daleth(\Psi(\hat{\mf t}_i),\Psi(\hat{\mf t}_j))\right) &= \Psi^{\star}\left(\left(\Omega_i^k\circ\phi^{-1}\right)\left(\Omega_j^l\circ\phi^{-1}\right)\ul\daleth(\Psi(\mf t_k),\Psi(\mf t_l)) + \left(\Omega_i^k\circ\phi^{-1}\right) \ul{M}_{kl} d\left(\Omega^l_j\circ \phi^{-1}\right)\right)\\
	& = \Omega_i^k\Omega_j^l\Psi^{\star}\left(\ul\daleth(\Psi(\mf t_k),\Psi(\mf t_l))\right) + \Omega_i^k {M}_{kl} d\Omega^l_j,
\end{align*}
where we used $\Psi^{\star} \ul{M}_{kl} = {M}_{kl}$ (by item (2) in Def. \ref{def_isometry}), $\Psi(\hat{\mf t}_i) = \Psi\left(\Omega^k_i\mf t_k\right)= \left(\Omega^k_i\circ\phi^{-1}\right) \Psi(\mf t_k)$ and that the pullback commutes with the exterior derivative. Since both sides in \eqref{condition3} transform in the same way, conditions \eqref{condition3} are invariant under change of basis of NPs.
\end{rmk}

\begin{rmk}
	The compatibility condition \eqref{conditionR}, namely $\phi^{\star}\left(\ul i^{\star}\bm{\mc{\ul R}}\right)=i^{\star}\bm{\mc R}$, played no essential role in \cite{Mio1} because in that paper we were interested in data satisfying the abstract constraint equations
	\begin{equation}
\label{constrainequations}
		\bm{\mc R} = \dfrac{2\Lambda}{m-1}\bg \quad \text{and}\quad \bm{\mc{\ul{R}}}=\dfrac{2\Lambda}{m-1}\ul\bg,
	\end{equation}
so \eqref{conditionR} was automatically fulfilled. When \eqref{constrainequations} do not hold (e.g. when the field equations are not the Einstein vacuum field equations or no equations whatsoever are imposed) adding this condition is essential. An explicit case is Theorem \ref{teo_suffi} below where we prove that given double null data $\{\mc D,\mc{\ul D},\mu\}$, there always exists a spacetime in which $\{\mc D,\mc{\ul D},\mu\}$ can be embedded. It turns out that without the condition $\phi^{\star}\left(\ul i^{\star}\bm{\mc{\ul R}}\right)=i^{\star}\bm{\mc R}$ the result would not be true.
\end{rmk}

DND has gauge freedom on each component linked by the behaviour of $\mu$ as described in Prop. \ref{proptransPsi}. Thus, we put forward the following definition.

\begin{defi}
	\label{defi_gaugeDND}
	Let $\{\mc D,\mc{\ul D},\mu\}$ be DND and $z\in\mc F^{\star}(\mc H)$, $\ul z\in\mc F^{\star}(\mc{\ul H})$, $\zeta\in\X(\mc H)$ and $\ul\zeta\in\X(\mc{\ul H})$. The transformed double null data is given by $\mc G\left(\{\mc D,\mc{\ul D},\mu\}\right)\d \{\mc D',\mc{\ul D}',\mu'\}$, where $\mc D'$ and $\mc{\ul D}'$ are the transformed CHD in the sense of Definition \ref{defi_gauge} and
	\begin{equation}
		\mu' \d z^{-1}\ul z^{-1} \mu.
	\end{equation}
\end{defi}

This definition guarantees that when $\{\mc D,\mc{\ul D},\mu\}$ is double null data, then $\mc G\left(\{\mc D,\mc{\ul D},\mu\}\right)$ is double null data too. This is a straightforward consequence of the gauge covariance of the compatibility conditions (see Remark \ref{remark_DNDgaugeinvariant}). In order to connect the abstract definition of double null data with the geometric idea of two null and transverse hypersurfaces, it is necessary to extend the notion of embeddedness to the context of double null data.

\begin{defi}
	\label{def_embDND}
	Let $\{\mc D,\mc{\ul D},\mu\}$ be DND and $(\mc M,g)$ a spacetime. We say that $\{\mc D,\mc{\ul D},\mu\}$ is embedded double null data on $(\mc M,g)$ with riggings $\xi,\ul\xi$ and embeddings $\Phi,\ul\Phi$, respectively, provided that
	\begin{enumerate}
		\item $\mc D$ (resp. $\mc{\ul D}$) is embedded in $(\mc M,g)$ with embedding $\Phi$ (resp. $\ul\Phi$) and rigging $\xi$ (resp. $\ul\xi$) in the sense of Def. \ref{defi_embedded} and $\Phi(\mc{S})=\ul\Phi(\mc{\ul S})\eqqcolon S$.
		\item $\mu(p)=g(\nu,\ul\nu)|_{\ul\Phi(p)}$ for all $p\in\mc{\ul S}$, where $\nu=\Phi_{\star}n$ and $\ul\nu = \ul\Phi_{\star}\ul n$.
	\end{enumerate}
\end{defi}

Definitions \ref{defi_embedded}, \ref{defi_DND} and \ref{def_embDND} establishes our intended correspondence between embedded double null data and two transverse, null hypersurfaces. Moreover, since the quantity $\ul{\mc A}\left(\Psi(n,0),(\ul n,0)\right)$ is assumed to be negative when $n$ and $\ul n$ point into the interior of $\mc H$ and $\mc{\ul H}$, respectively, the two hypersurfaces have the same time-orientation.\\

Let us summarize what we have done so far. In Definition \ref{defi_DND} we have introduced a completely abstract object called double null data that captures the idea of two null and transverse hypersurfaces but without seeing them as embedded in any ambient spacetime. This object must satisfy certain compatibility conditions at the ``corner'', which are clearly necessary for any DND to be able to be embedded. These conditions have been written in a fully gauge-covariant way thanks to the notion of normal pair. They are also independent of the basis of NPs. In the following remark we connect the new completely general definition of double null data with the previous one given in Def. 7.5 of \cite{Mio1}, where the compatibility equations were written in a particular gauge assuming strong conditions at the boundary as well as a very particular basis of normal pairs.

\begin{rmk}
	\label{remark2}
Using Lemma \ref{lema_escala} and Remark \ref{remark}, the compatibility conditions \eqref{condition1} can be written in the basis of normal pairs $\{\mf t_n =(n,0), \mf t_{\theta}=(\theta,1)\}$ as 
\begin{align}
\mu\Psi^{\star} \left(\ul{\bm\Upsilon} + \ul{\bm\Theta}\right) &= \bm\chi,\label{condition11}\\
\mu^{-1}\Psi^{\star}\ul{\bm\chi}&= \bm\Upsilon+\bm\Theta.\label{condition22}
\end{align}
In addition, using again Remark \ref{remark} and Corollary \ref{transdaleth}, \eqref{condition3} can be written as $$\daleth(\mf t_{\theta},\mf t_n) = \left(\Psi^{\star}\ul\daleth\right)\left(\Psi(\mf t_{\theta}),\Psi(\mf t_n)\right) = \left(\Psi^{\star}\ul\daleth\right)\left(\mu^{-1} \mf{\ul t}_n, \mu \mf{\ul t}_{\theta}\right) = \Psi^{\star}\bm{\ul\tau} +d(\log|\mu|),$$ so employing Prop. \ref{prop_daleths} it yields
\begin{equation}
\label{condition33}
\bm\tau + \Psi^{\star}\bm{\ul\tau} = -d(\log|\mu|).
\end{equation}
Given $\mc D$ and $\ul{\mc D}$ characteristic hypersurface data, there exists a class of gauges in which $\bm\ell_{\para}=0$, $\ell^{(2)}=0$ on $\mc{S}$ and $\bm{\ul\ell}_{\para}=0$, $\ul\ell^{(2)}=0$ on $\mc{\ul S}$. The existence of these class of gauges was established in \cite[Lemma 7.2]{Mio1} where it was also proved that this family of gauges was parametrized by pairs $(z,\zeta)$ and $(\ul z,\ul \zeta)$ satisfying $\zeta|_{\mc{S}}=0$ and $\ul\zeta|_{\mc{\ul S}}=0$. Particularizing equations \eqref{condition11}, \eqref{condition22} and \eqref{condition33} to this class of gauges, it yields
\begin{align}
\mu \ul\bY(X,Y) &= \bK(X,Y),\label{conditionnew1}\\
\mu^{-1} \ul\bK(X,Y) &= \bY(X,Y),\label{conditionnew2}\\
\bPi(X,n) + \ul\bPi(X,\ul n) & = -X(\log|\mu|)\label{conditionnew3}
\end{align} 
for every $X,Y\in\X(\mc{S})$ (here we omit the pushforward $\phi_{\star}$ for simplicity). These equations are the same as the compatibility conditions (122)-(124) in \cite{Mio1}. We conclude that Def. \ref{defi_DND} generalizes our previous definition of double null data in a fully general gauge and in any basis of normal pairs.
\end{rmk}

So far it is clear that the compatibility conditions \eqref{conditionh}-\eqref{conditionR} are necessary for a double null data to be embeddable in some spacetime. The rest of this section is devoted to show that these equations are not only necessary but also sufficient, i.e., that if \eqref{conditionh}-\eqref{conditionR} hold, then there exists a spacetime in which the double null data can be embedded. Here we are not interested in solving any spacetime field equations, we simply want to find that there always exists a spacetime where the data can be embedded, with the aim of showing that we have not forgotten any additional restriction on the data that might have been necessary.\\

The construction of the spacetime will be based on the harmonic gauge. In Theorem \ref{teo_HG} we have already recalled the result in \cite{Mio1} that guarantees the existence of a harmonic gauge associated to a set of $m$ functionally independent functions on $\mc H$. We now choose functions $\{x^{\ul a}\}=\{\ul u,x^A\}$ on $\mc H$ and $\{\ul x^{\ul a}\}=\{u,\ul x^A\}$ on $\mc{\ul H}$ with the following properties:
\begin{equation}
\label{hola}
\hskip -6mm \left.\begin{array}{ll}
(i) &\hskip -2mm n(\ul u)\neq 0,\, \ul u|_{\mc{S}}=0,\, n(x^A) = 0 \textup{ and } \{x^A\}$ \textup{is a local coordinate system on} $\mc{S}\\

(ii) &\hskip -2mm \ul n(u)\neq 0,\, u|_{\mc{\ul S}}=0,\, \ul n(\ul x^A) = 0 \textup{ and } \{\ul x^A\} \textup{ is a local coordinate system on } \mc{\ul S}\\

(iii) &\hskip -2mm \ul x^A\circ \phi = x^A
\end{array}\hskip -2mm\right\rbrace .
\end{equation}

Lemma \ref{HG2} guarantees the existence of a harmonic gauge associated to the functions $\{\ul u,x^A\}$ (resp. $\{u,\ul x^A\}$) on $\mc H$ (resp. $\mc{\ul H}$) satisfying $\bm\ell_{\para}=0$, $\ell^{(2)}=0$ on $\mc{S}$ and $\bm{\ul\ell}_{\para}=0$, $\ul\ell^{(2)}=0$ on $\mc{\ul S}$. Moreover, the residual gauge freedom is parametrized by pairs $(z,\ul z)\in \mc F^{\star}(\mc{S})\times \mc F^{\star}(\mc{S})$. One can exploit this freedom to fix the value of the functions $n(\ul u)$ and $\ul n(u)$ at $\mc{S}$ and $\mc{\ul S}$, respectively.

\begin{lema}
	\label{lema_compHG}
Let $\{\mc D,\mc{\ul D},\mu\}$ be DND and consider two set of independent functions $\{\ul u, x^{A}\}$ and $\{u,\ul x^{A}\}$ satisfying conditions \eqref{hola}. Then there exists a unique harmonic gauge w.r.t $\{\ul u,x^{A}\}$ and $\{u,\ul x^{A}\}$ in $\mc D$ and $\ul{\mc D}$, respectively, in which $\bm\ell_{\para}=0$, $\ell^{(2)}=0$, $n(\ul u)=\mu$ on $\mc{S}$ and $\bm{\ul\ell}_{\para}=0$, $\ul\ell^{(2)}=0$, $\ul n(u)=\mu$ on $\mc{\ul S}$.
\begin{proof}
The transformation law of the functions $n(\ul u)$ and $\ul n(u)$ follow from that of $n$ (see \eqref{transn}), $n'(\ul u) = z^{-1} n(\ul u)$ and $\ul n'(u)=\ul z^{-1}\ul n(u)$. Recalling the transformation of $\mu$ in \eqref{transmu}, namely $\mu' = z^{-1}\ul z^{-1} \mu$, one can choose $\ul z=\mu n(\ul u)^{-1}$ and $z=\mu \ul n(u)^{-1}$ so that $\mu' = n'(\ul u) = \ul n'(u)$.
\end{proof}
\end{lema}

Applying Proposition \ref{propHG3} it follows that when the DND is written in the unique gauge defined in the previous lemma, additional relations between the data at the boundary appear. 

\begin{prop}
	\label{prop_compatible3}
	Let $\{\mc D,\mc{\ul D},\mu\}$ be DND written in the harmonic gauge of Lemma \ref{lema_compHG} w.r.t $\{\ul u, x^{A}\}$ and $\{u,\ul x^{A}\}$. Then for every $X\in\X(\mc{S})$ the following relations hold at $\mc{S}$ (we omit the pushforward $\phi_{\star}$ for simplicity)
	\begin{align}
		2\bY(n,n)&=\mu\ul n\big(\ul\ell^{(2)}\big) \label{omega},\\
		2\ul \bY(\ul n,\ul n) &=\mu n\left(\ell^{(2)}\right) \label{ulomega},\\
		2\bPi(X, n) &=\left(\lie_{\ul n}\bm{\ul\ell}\right)(X) -X\left(\log|\mu|\right) - \left(\lie_n\bm\ell\right)(X),\label{tau}\\
		2\ul\bPi(X,\ul n) &= \left(\lie_{n}\bm{\ell}\right)(X) -X\left(\log|\mu|\right) - \left(\lie_{\ul n}\bm{\ul\ell}\right)(X).\label{ultau}
	\end{align}
\begin{proof}
Firstly, from equation \eqref{eqHG1} it follows $n\left(\ell^{(2)}\right) = \tr_h\bm\Upsilon$ on $\mc{S}$, which together with \eqref{conditionnew2} and \eqref{combGamma} yields \eqref{ulomega}. Equation \eqref{omega} is analogous. Secondly, from \eqref{eqHG2} and $\ul x^A\circ \phi = x^A$, $$2\left(\lie_n\bm\ell + \bPi(\cdot,n)\right)(\grad_h x^A) = \square_h x^A = \square_{\ul h} \ul x^A = 2\left(\lie_{\ul n}\bm{\ul\ell} + \ul\bPi(\cdot,\ul n)\right)(\grad_{\ul h} \ul x^A).$$ Employing \eqref{conditionnew3}, equations \eqref{tau} and \eqref{ultau} follow at once.
\end{proof}
\end{prop}

In the following remark we compute the compatibility condition \eqref{conditionR} in the gauge defined in Lemma \ref{lema_compHG}, for which we first need to write the pullback of the tensor $\bm{\mc R}$ into a section. This has been computed in \cite{miguel3} in the case of general null hypersurface data. Their result is fully diffeomorphism covariant, the dependence on the tensor $\bY$ is explicit and is written without assuming any gauge condition. However, since we only need $\bm{\mc R}$ in a very specific gauge and to make this article as self-contained as possible, we redo this computation in Appendix \ref{appendix} assuming a gauge in which both $\ell^{(2)}$ and $\bm\ell_{\para}$ vanish at the section.

\begin{rmk}
In this remark we compute the compatibility condition $\ul i^{\star}\bm{\mc{\ul R}} = i^{\star}\bm{\mc R}$ on $\mc{S}$ in the gauge defined in Lemma \ref{lema_compHG} (we omit the pullback $\phi^{\star}$ for simplicity). In Lemma \ref{lemapendice} of Appendix \ref{appendix} we found that the pullback to $\mc{S}$ of the abstract constraint tensor $\bm{\mc R}$ as defined in \eqref{Rab} in a gauge in which $\bm\ell_{\para}=0$ and $\ell^{(2)}=0$ on $\mc{S}$ is
\begin{equation}
	\label{RAB2}
	\begin{aligned}
		\bm{\mc R}_{AB} & = R^h_{AB} + \left(2\bY(n,n) - \tr_h\bK\right) \bY_{AB} + \left(n\left(\ell^{(2)}\right) - \tr_h\bY\right)\bK_{AB}\\
		&\quad + 4h^{CD}\bK_{C(A}\bY_{B)D} + 2\nabla_{(A}^h\bm\tau_{B)} + 2\nabla_{(A}^h\left(\lie_n\bm\ell\right)_{B)}- 2\lie_n\bPi_{(AB)}- 2\bm\tau_A\bm\tau_B,
	\end{aligned}
\end{equation}
and analogously the pullback of $\bm{\mc{\ul R}}$ into $\mc{\ul S}$ in the same gauge is 
\begin{equation}
	\label{ulRAB2}
	\begin{aligned}
		\bm{\mc{\ul R}}_{AB} & = R^{\ul h}_{AB} + \left(2\ul\bY(\ul n,\ul n) - \tr_{\ul h}\ul\bK\right) \ul\bY_{AB} + \big(\ul n\big(\ul\ell^{(2)}\big) - \tr_{\ul h}\ul\bY\big)\ul\bK_{AB}\\
		&\quad + 4\ul h^{CD}\ul\bK_{C(A}\ul\bY_{B)D} + 2\nabla_{(A}^{\ul h}\bm{\ul\tau}_{B)} + 2\nabla_{(A}^{\ul h}\left(\lie_{\ul n}\bm{\ul\ell}\right)_{B)}- 2\lie_{\ul n}\ul\bPi_{(AB)}- 2\bm{\ul\tau}_A\bm{\ul\tau}_B.
	\end{aligned}
\end{equation}
From condition \eqref{conditionh}, namely that $h\st{\mc S}{=}\ul h$, it follows $ R^h_{AB} \st{\mc S}{=} R^{\ul h}_{AB}$. From \eqref{conditionnew1}-\eqref{conditionnew2}, the terms $\tr_h\bK\  \bY_{AB}$, $\tr_h\bY\ \bK_{AB}$ and $4h^{CD}\bK_{C(A}\bY_{B)D}$ from \eqref{RAB2} are equal to the terms $\tr_{\ul h}\ul\bY\ \ul\bK_{AB}$, $\tr_{\ul h}\ul\bK\ \ul\bY_{AB}$ and $4\ul h^{CD}\ul\bK_{C(A}\ul\bY_{B)D}$ from \eqref{ulRAB2}, respectively. Finally from condition \eqref{condition33} we can substitute $\bm{\ul\tau}$ in \eqref{ulRAB2} in terms of $\bm\tau$ and $d\log|\mu|$. Then, the compatibility condition $\bm{\mc R}_{AB} = \bm{\mc{\ul R}}_{AB}$ in the gauge in which $\bm\ell_{\para}=\ul{\bm\ell}_{\para}=0$ and $\ell^{(2)}=\ul\ell^{(2)}=0$ on $\mc{S}$ and $\mc{\ul S}$ can be written as
\begin{align}
\hskip -6mm \left(\lie_n\bPi\right)_{(AB)} - \left(\lie_{\ul n}\ul\bPi\right)_{(AB)} & = \left(\bY(n,n) - \dfrac{1}{2}\mu \ul n\big(\ul\ell^{(2)}\big)\right)\bY_{AB} + \left(\dfrac{1}{2} n\left(\ell^{(2)}\right) -\mu^{-1}\ul\bY(\ul n,\ul n)\right)\bK_{AB}\nonumber\\
&\quad  +2\nabla^h_{(A}\bm\tau_{B)} + \nabla^h_{(A}\left(\lie_n\bm\ell\right)_{B)} - \nabla^h_{(A}\left(\lie_{\ul n}\bm{\ul\ell}\right)_{B)} +\nabla^h_{(A}\nabla^h_{B)}\log|\mu|\label{LienPi}\\
&\quad + 2\bm\tau_{(A}\nabla^h_{B)}\log|\mu| + \nabla^h_A\log|\mu| \nabla^h_B\log|\mu|,\nonumber
\end{align}
We now restrict further the gauge so that we are in the harmonic gauge of Lemma \ref{lema_compHG}. Taking into account \eqref{omega}-\eqref{ultau} and the fact that $\bPi(X,n)=\bm\tau(X)$ for every $X\in\X(\mc{S})$, the first and second lines of the RHS of \eqref{LienPi} vanish. Replacing the term $2\bm\tau_A$ of the third line by $\left(\lie_{\ul n}\bm{\ul\ell}\right)_A - \nabla^h_A\log|\mu| - \left(\lie_n\bm\ell\right)_A$ (see \eqref{tau}), equation \eqref{LienPi} finally reads 
\begin{equation}
	\label{LiePicoor}
\left(\lie_n\bPi\right)_{(AB)} - \left(\lie_{\ul n}\bPi\right)_{(AB)} = \left(\lie_{\ul n}\bm{\ul\ell}\right)_{(A} \nabla^h_{B)}\log|\mu| - \left(\lie_{n}\bm{\ell}\right)_{(A} \nabla^h_{B)}\log|\mu|.
\end{equation}
\end{rmk}

We now have the necessary ingredients to show that \eqref{conditionh}-\eqref{conditionR} is everything one needs to make sure that double null data can be embedded in some spacetime, i.e., that the definition is complete and we are not missing any extra conditions.

\begin{teo}
	\label{teo_suffi}
Let $\{\mc D,\ul{\mc D},\mu\}$ be double null data. Then there exists a spacetime $(\mc M,g)$, embeddings $\Phi:\mc H\hookrightarrow\mc M$, $\ul \Phi:\mc{\ul H}\hookrightarrow\mc M$ and vector fields $\xi,\ul\xi$ along $\Phi(\mc H)$ and $\ul\Phi(\mc{\ul H})$, respectively, such that $\{\mc D,\ul{\mc D},\mu\}$ is embedded double null data in $(\mc M,g)$ with embeddings $\Phi$, $\ul\Phi$ and riggings $\xi,\ul\xi$, respectively.
\begin{proof}
Let $\{\mc D,\ul{\mc D},\mu\}$ be double null data, $\mc D=\{\mc H,\bg,\bm\ell,\ell^{(2)},\bY\}$ and $\mc{\ul D}=\{\mc{\ul H},\ul\bg,\bm{\ul\ell},\ul\ell^{(2)},\ul \bY\}$. Consider a set of independent functions $\{\ul u, x^A\}$ and $\{u,\ul x^A\}$ satisfying the conditions of Lemma \ref{lema_compHG}, namely \eqref{hola}. Henceforth the coordinates $u$ and $\ul u$ will be assumed to be $\ge 0$. We write the data in the gauge of Lemma \ref{lema_compHG} with respect to these functions. Consider the manifold $\real\times\real\times\mc{S}$ and use $u$ and $\ul u$ as the natural coordinates in the first and second factors, respectively. We shall work in the manifold with boundary $\mc M\d\{u,\ul u\ge 0\}\subset \real^2\times\mc{S}$. Any (local) coordinate system $\{x^A\}$ in $\mc{S}$ extends to a (local) coordinate system $\{u,\ul u,x^A\}$ in $\mc M$ such that $\{\ul u,x^A\}$ (resp. $\{u,x^A\}$) restricted to $\mc H$ (resp. $\mc{\ul H}$) are the given coordinates on $\mc H$ (resp. $\mc{\ul H}$), as well as $u|_{\mc H}=0$ and $\ul u|_{\mc H}=0$. We define the embeddings
\begin{equation}
\begin{array}{lllll}
	\Phi: \mc H&\to\mc M & \qquad&\ul\Phi: \mc{\ul H}&\to\mc M\\
	(\ul u, x^A)&\mapsto (u=0,\ul u, x^A) &\qquad & (u,\ul x^A)&\mapsto (u,\ul u=0, \ul x^A).
\end{array}
\end{equation}
Thus, $\Phi(\mc H)=\{u=0\}$ and $\ul \Phi(\mc{\ul H})=\{\ul u=0\}$. We denote by $S$ the intersection of $\Phi(\mc H)$ and $\ul \Phi(\mc{\ul H})$, namely $S \d \{u=\ul u=0\}\subset\mc M$. Let $\xi$ and $\ul\xi$ be defined by $\xi\d\partial_u$ and $\ul\xi\d\partial_{\ul u}$ in the coordinate system we have introduced. In these coordinates we also have $n = \lambda\partial_{\ul u}$ and $\ul n=\ul\lambda \partial_u$, where $\lambda\d n(\ul u)$ and $\ul\lambda\d \ul n(u)$. In order to prove the theorem we only need to construct a smooth metric $g$ on $\mc M$ inducing the given data on $\mc H\cup\ul{\mc H}$ (we do not write $\Phi$ or $\ul\Phi$ for simplicity) w.r.t the riggings $\xi$ and $\ul\xi$, i.e., 
\begin{equation}
	\label{ind1}
	\begin{array}{lllll}
	g_{\ul u\,\ul u}|_{\mc H} &= 0, \quad \hfill g_{u\, u}|_{\mc H} &= \ell^{(2)}, \hspace{2.5cm} \hfill  g_{\ul u\,\ul u}|_{\mc{\ul H}}&=\ul\ell^{(2)}, \quad g_{u\, u}|_{\mc{\ul H}} &= 0,\\
	g_{\ul u\, A}|_{\mc H} &= 0, \hfill g_{u\, A}|_{\mc H} &= \bm\ell_A, \hspace{2.5cm} \hfill g_{\ul u\, A}|_{\mc{\ul H}}&=\bm{\ul\ell}_A, \quad  g_{u\, A}|_{\mc{\ul H}}&=0,\\
	& \hfill g_{AB}|_{\mc H} &= \bg_{AB}, \hspace{2.5cm} \hfill g_{AB}|_{\mc{\ul H}} &= \ul \bg_{AB},& \\
	&\hfill g_{u \, \ul u}|_{\mc H} &= \lambda^{-1}, \hfill g_{u\, \ul u}|_{\mc{\ul H}} &= \ul\lambda^{-1},
	\end{array}
\end{equation}
\begin{equation}
	\label{ind1.5}
	g_{u\, \ul u}|_{S} = \mu^{-1},
\end{equation}
and
\begin{equation}
	\label{ind2}
	\dfrac{1}{2}\left(\lie_{\xi} g\right)_{ab} = \bY_{ab},\qquad \dfrac{1}{2}\big(\lie_{\ul\xi} g\big)_{ab} = \ul\bY_{ab},
\end{equation}
where $\bm\ell_A\d \bm\ell(\partial_{x^A})$, $\bg_{AB}\d \bg(\partial_{x^A},\partial_{x^B})$ and similarly on $\mc{\ul H}$. The conditions of the first line of \eqref{ind1} follow from the fact that $\bg(n,\cdot)=0$ and $\Phi^{\star}\left(g(\xi,\xi)\right)=\ell^{(2)}$ (see \eqref{embedded_equations}). The second line also follows from $\bg(n,\cdot)=0$ as well as from $\Phi^{\star}\left(g(\xi,\cdot)\right)=\bm\ell$. The third line follows directly from $\Phi^{\star} g = \bg$. The fourth line follows from $1=g(\xi,\nu)=g(\partial_u,\lambda\partial_{\ul u})=\lambda g_{u\,\ul u}$ and its underlined version. These four lines constitute the metric part of the data. Condition \eqref{ind1.5} follows from $\mu=g(\nu,\ul \nu)|_{\mc S} = \lambda\ul\lambda g_{u\,\ul u}$ and the fact that $\lambda=\ul\lambda=\mu$ on $S$. Finally, conditions \eqref{ind2} guarantee that the metric $g$ also induces the $\bY$ tensor (see \eqref{Yembedded}).\\

In order to construct such $g$, our strategy is to extend the components of the hypersurface data tensors in these coordinates to all $\mc M$ and define the components of the metric $g$ in such a way that it induces the given data on $\mc H\cup\ul{\mc H}$. Thus, we introduce the notation $f^{\mc H}$ to denote the extension of the function $f\in \mc F(\mc H)$ off $\mc H$ satisfying $\xi\left(f\right)=0$. We define $f^{\mc{\ul H}}$ analogously, i.e., by extending the function $f\in\mc F(\mc{\ul H})$ by means of $\ul\xi(f)=0$. Moreover, $f^{S}$ will denote the extension of $f\in\mc F(S)$ off $S$ satisfying $\xi\left(f\right) = \ul\xi\left(f\right)=0$. Then we define the components of $g$ in this coordinate system as follows.
\begin{itemize}
	\item Component $g_{\ul u\, \ul u}$: Let $g_{\ul u\, \ul u}$ be defined on $\mc M$ by $g_{\ul u\,\ul u}\d (\ul\ell^{(2)})^{\mc{\ul H}}+2(\bY_{\ul u\, \ul u}^{\mc{H}}-\bY_{\ul u\, \ul u}^{S})u$. Since $\bY^{\mc H}_{\ul u\, \ul u} = \bY^{S}_{\ul u\, \ul u}$ on $\mc{\ul H}$ we have $g_{\ul u\, \ul u}|_{\mc{\ul H}}=\ul\ell^{(2)}$. From $\ul\ell^{(2)}\st{S}{=}0$ its extension $(\ul\ell^{(2)})^{\mc{\ul H}}$ vanishes on $\mc H$ and hence $g_{\ul u\, \ul u}|_{\mc{H}}=0$. Concerning the transverse derivative, we first note that for any function $f\in\mc F(\mc H)$ it holds $\partial_{\ul u} (f^{\mc{H}}) = (\partial_{\ul u} f)^{\mc H}$ and similarly $\partial_u (f^{\mc{\ul H}}) = (\partial_u f)^{\mc{\ul H}}$ for any function $f\in\mc F(\mc{\ul H})$. Indeed, $\partial_u \left(\partial_{\ul u} (f^{\mc{H}})\right)= \partial_{\ul u} \left(\partial_u (f^{\mc{H}})\right) = 0$ and $\partial_u \left((\partial_{\ul u} f)^{\mc H}\right)=0$ by construction, so the function $\partial_{\ul u} (f^{\mc{H}}) - (\partial_{\ul u} f)^{\mc H}$ is constant along each integral curve of $\partial_u$, and since on $\mc H$ $\partial_{\ul u} (f^{\mc{H}}) = (\partial_{\ul u} f)^{\mc H}=\partial_{\ul u} f$, we conclude that $\partial_{\ul u} (f^{\mc{H}}) = (\partial_{\ul u} f)^{\mc H}$ (and similarly $\partial_u (f^{\mc{\ul H}}) = (\partial_u f)^{\mc{\ul H}}$ on $\mc M$). Now, condition \eqref{omega} together with $n = \lambda\partial_{\ul u}$, $\ul n=\ul\lambda \partial_u$ and $\lambda=\ul\lambda$ on $S$ gives $\partial_u \ul\ell^{(2)} \st{S}{=} 2 \bY_{\ul u\,\ul u}$. Therefore, all along $\mc H$ their extensions agree, $(\partial_u\ul\ell^{(2)})^{\mc{\ul H}} = \partial_u \big((\ul\ell^{(2)})^{\ul{\mc H}}\big) = 2\bY^{S}_{\ul u\, \ul u}$. Consequently,	
	$$\dfrac{1}{2}\left(\lie_{\xi} g\right)_{\ul u\, \ul u} = \dfrac{1}{2}\partial_u g_{\ul u\, \ul u} = \dfrac{1}{2}\partial_u \big((\ul\ell^{(2)})^{\mc{\ul H}}\big) + \bY_{\ul u\, \ul u}^{\mc{H}} - \bY_{\ul u\, \ul u}^{S} \st{\mc H}{=} \bY_{\ul u\, \ul u}.$$ 
	\item Component $g_{u\, u}$: Analogously, we define $g_{u\, u}\d (\ell^{(2)})^{\mc H} + 2\left(\ul\bY_{u\, u}^{\mc{\ul H}}- \ul\bY_{u\, u}^{S}\right)\ul u$. By symmetry of the construction this also induces the given data on $\mc H$ and $\mc{\ul H}$.

	\item Component $g_{u\, \ul u}$: Let $g_{u\, \ul u}$ be defined on $\mc M$ by $g_{u\, \ul u} \d \left(\lambda^{-1}\right)^{\mc H} + \left(\ul\lambda^{-1}\right)^{\mc{\ul H}} - \left(\mu^{-1}\right)^{S}$. Since $\mu=\lambda=\ul\lambda$ on $\mc S$, $g_{u\, \ul u}|_{\mc H} = \lambda^{-1}$ and $g_{u\, \ul u}|_{\mc{\ul H}} = \ul\lambda^{-1}$ (and they match on $S$ and fulfil condition \eqref{ind1.5}).

% Me estoy haciendo un puto lio de cuidado con las putas u y u barra
	
	\item Components $g_{\ul u\, A}$: Let $g_{\ul u\, A}\d \bm{\ul\ell}_A^{\ul{\mc H}} + 2(\bY_{\ul u\, A}^{\mc H} - \bY_{\ul u\, A}^{S})u$ on $\mc M$. Since $\bY_{\ul u\, A}^{\mc H} =\bY_{\ul u\, A}^{S}$ on $\mc{\ul H}$ we have $g_{\ul u\, A}|_{\mc{\ul H}}= \bm{\ul\ell}_A$. From $\bm{\ul\ell}_A=0$ on $S$, its extension $\bm{\ul\ell}_A^{\mc{\ul H}}$ also vanishes on $\mc H$, and then $g_{\ul u\, A}|_{\mc{H}}=0$. In order to see that these $g_{\ul u\, A}$ induce the corresponding components of the $\bY$ tensor we start by writing equation \eqref{tau} in the coordinate system $\{u,\ul u,x^A\}$, i.e., taking $X=\partial_{x^A}$, $n=\lambda\partial_{\ul u}$ and $\ul n=\ul\lambda\partial_u$. Using $\mu\st{S}{=}\ul\lambda$,
\begin{equation}
\label{auxx1}
\begin{aligned}
2\lambda \bPi_{A\,\ul u} & \st{S}{=}\ul\lambda \partial_u \bm{\ul\ell}_A + \ul{\bm\ell}(\partial_u)\partial_{x^A}(\ul\lambda) - \partial_{x^A} \log|\ul\lambda| - \lambda \partial_{\ul u}\bm\ell_A - \bm\ell(\partial_{\ul u})\partial_{x^A}(\lambda)\\
& \st{S}{=}\ul\lambda \partial_u \bm{\ul\ell}_A + \ul\lambda^{-1}\partial_{x^A}(\ul\lambda) - \partial_{x^A} \log|\ul\lambda| - \lambda \partial_{\ul u}\bm\ell_A - \lambda^{-1}\partial_{x^A}(\lambda)\\
& \st{S}{=} \ul\lambda \partial_u \bm{\ul\ell}_A - \partial_{x^A} \log|\ul\lambda| - \lambda \partial_{\ul u}\bm\ell_A,
\end{aligned}
\end{equation}
where in the first line we used the well-known formula $\lie_{fX}\bm\omega = f\lie_X\bm\omega + \bm\omega(X) df$ valid for any function $f$, vector $X$ and one-form $\bm\omega$, in the second line we used $\bm\ell(n)=\lambda\bm\ell(\partial_{\ul u})=1$ and thus $\bm\ell(\partial_{\ul u}) =\lambda^{-1}$ (and its underlined version) and in the third line $\ul\lambda\partial_{x^A}(\ul\lambda) \st{S}{=} \lambda\partial_{x^A}(\lambda)$ since $\lambda=\ul\lambda$ on $S$. The value of $\bPi_{A\,\ul u}$ can be computed from \eqref{defPi} and \eqref{def_F}, namely
	 \begin{equation}
	 	\label{auxx2}
	 	2\bPi_{A\,\ul u} = 2\bY_{A\,\ul u} + 2\bF_{A\,\ul u} = 2\bY_{A\,\ul u} + \partial_{x^A} \lambda^{-1} - \partial_{\ul u}\bm\ell_A,
	 \end{equation}
where again we used $\bm\ell_{\ul u} = \lambda^{-1}$. Inserting \eqref{auxx2} evaluated at $S$ into \eqref{auxx1} and taking again into account that $\lambda=\ul\lambda$ on $\mc S$, it yields $2 \bY_{\ul u\, A}=\partial_{u} \bm{\ul\ell}_A$ on $S$ and therefore $2 \bY_{\ul u\, A}^{S}=\left(\partial_{u} \bm{\ul\ell}_A\right)^{\mc{\ul H}}=\partial_{u}\big(\bm{\ul\ell}_A^{\mc{\ul H}}\big)$ on $\mc H$. Hence, $$\dfrac{1}{2}\left(\lie_{\xi} g\right)_{\ul u\, A} = \dfrac{1}{2} \partial_{u} g_{\ul u\, A} = \dfrac{1}{2}\partial_u\big(\bm{\ul\ell}_A^{\mc{\ul H}}\big) + {\bY}_{\ul u\, A}^{\mc H} - {\bY}_{\ul u\, A}^{S} \st{\mc H}{=} {\bY}_{\ul u\, A}.$$

	\item Components $g_{ u\, A}$: Analogously, we define $g_{ u\, A}\d \bm{\ell}_A^{\mc H} + 2(\ul\bY_{ u\, A}^{\mc{\ul H}} - \bY_{ u\, A}^{S})\ul u$ on $\mc M$, which by symmetry also induces the given data on $\mc H$ and $\mc{\ul H}$.
	
% - \left((\partial_{\ul u}\bY_{AB}^{\mc{H}})^{\mc S}+(\partial_u\ul\bY^{\mc{\ul H}}_{AB})^{\mc S}\right)u\ul u

	\item Components $g_{AB}$: Let $h$ be the induced metric on $\mc S$. We define the functions $g_{AB}$ on $\mc M$ by means of 
	\begin{align*}
		g_{AB} \d \bg_{AB}^{\mc H} + \ul\bg_{AB}^{\mc{\ul H}} - h_{AB}^{S} + 2\left(\bY_{AB}^{\mc H}-\bY_{AB}^{S}\right)u + 2\big(\ul \bY_{AB}^{\mc{\ul H}}-\ul \bY_{AB}^{S}\big)\ul u -2(\partial_{u} \ul\bY_{AB})^{S} u\ul u.
	\end{align*} 
Since on $\mc H$ $\ul\bg_{AB}^{\ul{\mc H}} = h_{AB}^{S}$ and $\ul\bY_{AB}^{\ul{\mc H}}=\ul\bY_{AB}^{S}$, and on $\mc{\ul H}$ $\bg_{AB}^{\mc H} = h_{AB}^{S}$ and $\bY_{AB}^{\mc H}=\bY_{AB}^{S}$, we have $g_{AB}|_{\mc H}= \bg_{AB}$ and $g_{AB}|_{\mc{\ul H}}= \ul\bg_{AB}$. Moreover, from \eqref{defK} and $\ul n=\ul\lambda\partial_u$, we have $\partial_u \ul\bg_{AB} = 2\ul\lambda^{-1}\ul\bK_{AB}$ on $\mc{\ul H}$, so in particular $\partial_u \ul\bg_{AB} = 2\ul\lambda^{-1}\ul\bK_{AB}$ on $S$. Using $\ul\lambda=\mu$ and $\mu^{-1}\ul{\bK}_{AB} = \bY_{AB}$ on $\mc S$ (see \eqref{conditionnew2}) it follows that $(\partial_u \ul\bg_{AB})^{\mc{\ul H}}= \partial_u\big( \ul\bg_{AB}^{\mc{\ul H}}\big) = 2\bY_{AB}^{S}$ on $\mc H$, and thus
\begin{align*}
\dfrac{1}{2}\left(\lie_{\xi} g\right)_{AB} &= \dfrac{1}{2}\partial_u g_{AB}\\
& = \dfrac{1}{2}\partial_u(\ul\bg_{AB}^{\mc{\ul H}}) + \bY_{AB}^{\mc H} - \bY_{AB}^{S} +\partial_u(\ul\bY_{AB}^{\mc{\ul H}})\ul u- \left(\partial_u\ul\bY_{AB}\right)^{S}\ul u\\
& \st{\mc H}{=} \dfrac{1}{2}\partial_u(\ul\bg_{AB})^{\mc{\ul H}} + \bY_{AB}^{\mc H} - \bY_{AB}^{S}\\
& \st{\mc H}{=} \bY_{AB},
\end{align*}
where in the third equality we used $\partial_u\big(\ul\bY_{AB}^{\mc{\ul H}}\big)=\left(\partial_u\ul\bY_{AB}\right)^{S}$ on $\mc H$. Before computing $\lie_{\ul\xi}g$ on $\ul{\mc H}$ we need to write equation \eqref{LiePicoor} in the coordinate system $\{u,\ul u, x^A\}$. Since $[n,\partial_{x^A}] = [\lambda\partial_{\ul u},\partial_{x^A}] = - \partial_{x^A}\lambda \ \partial_{\ul u}$ and $\lambda=\mu$ on $S$, $$\left(\lie_n\bPi\right)_{AB} \st{S}{=} n\left(\bPi_{AB}\right) + \partial_{x^A}(\log|\mu| )\bPi\left(n,\partial_{x^B}\right)+ \partial_{x^B}(\log|\mu|)\bPi\left(\partial_{x^A}, n\right).$$ Using \eqref{cartan} and the fact that $\bF$ is antisymmetric,
\begin{align*}
\left(\lie_n\bPi\right)_{AB} +\left(\lie_n\bPi\right)_{BA} &\st{S}{=} 2 n \left(\bY_{AB}\right) + \left(2\bPi(\partial_{x^B},n) + \left(\lie_n\bm\ell\right)(\partial_{x^B})\right)\partial_{x^A}(\log|\mu|)  \\
&\quad +\left(2\bPi(\partial_{x^A},n)+ \left(\lie_n\bm\ell\right)(\partial_{x^A})\right)\partial_{x^B}(\log|\mu|)\\
& \st{S}{=} 2 \lambda\partial_{\ul u} \left(\bY_{AB}\right) + \left(\lie_{\ul n}\bm{\ul\ell}\right)(\partial_{x^B}) \partial_{x^A}(\log|\mu|) + \left(\lie_{\ul n}\bm{\ul\ell}\right)(\partial_{x^A}) \partial_{x^B}(\log|\mu|)\\
&\quad  - 2\partial_{x^A}(\log|\mu|)\partial_{x^B}(\log|\mu|),
\end{align*}
where in the second equality we used \eqref{tau}. Then, equation \eqref{LiePicoor} in this coordinate system becomes simply
\begin{equation}
	\label{YYeq}
\partial_{\ul u}\left(\bY_{AB}\right)\st{S}{=} \partial_{u}\left(\ul\bY_{AB}\right),
\end{equation}
and then the quantity $\lie_{\ul\xi}g$ on $\ul{\mc H}$ is finally given by
\begin{align*}
	\dfrac{1}{2}\big(\lie_{\ul\xi} g\big)_{AB} &= \dfrac{1}{2}\partial_{\ul u} g_{AB}\\
	& = \dfrac{1}{2}\partial_{\ul u}\left(\bg_{AB}^{\mc{H}}\right) + \partial_{\ul u}\left(\bY^{\mc H}_{AB}\right) u + \ul\bY_{AB}^{\mc{\ul H}} - \ul\bY_{AB}^{S} - \left(\partial_u\ul\bY_{AB}\right)^{S} u\\
	& \st{\st{\mc{\ul H}}{}}{=} \dfrac{1}{2}\partial_u\left(\bg_{AB}\right)^{\mc{H}} + \ul\bY_{AB}^{\mc{\ul H}} - \ul\bY_{AB}^{S}\\
	& \st{\st{\mc{\ul H}}{}}{=} \ul\bY_{AB},
\end{align*}
where in the third line we used that on $\mc{\ul H}$ $\partial_{\ul u}\left(\bY^{\mc H}_{AB}\right) =(\partial_{\ul u}\bY_{AB})^{S} = \left(\partial_u\ul\bY_{AB}\right)^{S}$ (the second equality following from \eqref{YYeq}), and in the fourth line that $(\partial_{\ul u}\bg_{AB})^{\mc{H}}= \partial_{\ul u}\left(\bg_{AB}^{\mc{H}}\right) = 2\ul\bY_{AB}^{S}$ on $\mc{\ul H}$.
\end{itemize}
Given that $g_{\mu\nu}$ fulfils all conditions \eqref{ind1}-\eqref{ind2}, we conclude that $\{\mc D,\ul{\mc D},\mu\}$ is embedded double null data in $(\mc M,g)$ with embeddings $\Phi$, $\ul\Phi$ and riggings $\xi,\ul\xi$, respectively.
\end{proof}
\end{teo}
%$\ul\bY_{AB}^{\mc{\ul H}}=\ul\bY_{AB}^{S}$

The previous theorem shows that any double null data can be embedded in some spacetime. It then arises the natural question of whether it can be also embedded in a spacetime solution of the Einstein equations. By \eqref{Ricciembedded}, if $\{\mc D,\ul{\mc D},\mu\}$ is embedded DND on an $(m+1)$-dimensional spacetime $(\mc M,g)$ solution of the $\Lambda$-vacuum equations, namely $$\ric = \dfrac{2\Lambda}{m-1}g,$$ then it must satisfy
\begin{equation*}
	\bm{\mc R} = \dfrac{2\Lambda}{m-1}\bg \quad \text{and}\quad \bm{\mc{\ul{R}}}=\dfrac{2\Lambda}{m-1}\ul\bg,
\end{equation*}
where $\bm{\mc R}$ is the abstract tensor defined in \eqref{Rab} (and analogously for $\bm{\ul{\mc{R}}}$). Thus, these restrictions are necessary for $\{\mc D,\ul{\mc D},\mu\}$ to be embedded in a spacetime solution of the $\Lambda$-vacuum Einstein equations. The main result in \cite{Mio1} proves that they are also sufficient, as we summarize next.

\begin{defi}
	\label{development}
Let $\{\mc D,\mc{\ul D},\mu\}$ be double null data of dimension $m>1$. We say that a Lorentzian manifold $(\mc M,g)$ is a \textbf{development} of $\{\mc D,\mc{\ul D},\mu\}$ provided there exist embeddings $\Phi$, $\ul\Phi$ and riggings $\xi$, $\ul\xi$ such that $\{\mc D,\mc{\ul D},\mu\}$ is embedded DND in $(\mc M,g)$ with embeddings $\Phi$, $\ul\Phi$ and riggings $\xi$, $\ul\xi$ in the sense of Def. \ref{def_embDND} and $\Phi(\mc H)\cup \ul\Phi(\ul{\mc H})=\partial \mc M$.
\end{defi}

With this definition we can restate Theorem 7.15 of \cite{Mio1} in the following way.

\begin{teo}
	\label{main}
	Let $\{\mc D,\mc{\ul D},\mu\}$ be double null data of dimension $m> 1$ as defined in Def. \ref{defi_DND} satisfying the abstract constraint equations 
	\begin{equation}
		\label{constraintsL}
		\bm{\mc R} = \dfrac{2\Lambda}{m-1}\bg \quad \text{and}\quad \bm{\mc{\ul{R}}}=\dfrac{2\Lambda}{m-1}\ul\bg,
	\end{equation}
	where $\bm{\mc R}$ is defined in \eqref{Rab}, $\ul{\bm{\mc R}}$ is its underlined version and $\Lambda\in\real$. Then there exists a development $(\mc M,g)$ of $\{\mc D,\mc{\ul D},\mu\}$ (possibly restricted if necessary) solution of the $\Lambda$-vacuum Einstein equations. Moreover, for any two such developments $(\mc M,g)$ and $(\mc{\wh M},\wh g)$, there exist neighbourhoods of $\mc H\cup\mc{\ul H}$, $\mc U\subseteq\mc M$ and $\wh{\mc U}\subseteq\mc{\wh M}$, and a diffeomorphism $\varphi: \mc U\to \wh{\mc U}$ such that $\varphi^{\star}\wh g=g$.
\end{teo}

\begin{rmk}
Theorems \ref{teo_suffi} and \ref{main} establish a very clear hierarchy between the compatibility conditions and the constraint equations. The former are the necessary and sufficient conditions for a DND to be able to be embedded in some spacetime, whereas the later are necessary and sufficient for the DND to be embedded in a spacetime solution of the Einstein field equations.
\end{rmk}

\section{Isometry between Double Null Data}
\label{sec_isometry}

Given two double null data, there arises the natural question of under which conditions their developments are the same (up to isometry). In this section we establish the necessary and sufficient conditions for two double null data to define two isometric spacetimes. We start with a definition to fix some notation.

\begin{defi}
	\label{pullback1}
Let $\mc D=\{\mc H,\bg,\bm\ell,\ell^{(2)},\bY\}$ be hypersurface data and $\psi:\wh{\mc{H}}\to\mc H$ a diffeomorphism. We define the pull-back hypersurface data $\psi^{\star}\mc D$ by $$\psi^{\star}\mc D \d \left\{\wh{\mc{H}},\wh\bg\d\psi^{\star}\bg,\wh{\bm\ell}\d\psi^{\star}\bm\ell,\wh{\ell}^{(2)}\d\psi^{\star}\ell^{(2)},\wh{\bY}\d\psi^{\star}\bY\right\}.$$
\end{defi}

From Def. \ref{defi_hypersurfacedata} and the fact that $\psi$ is a diffeomorphism, it follows that $\psi^{\star}\mc D$ is still hypersurface data. Moreover, from \eqref{gamman}-\eqref{gammaP} having a unique solution for $\{P,n,n^{(2)}\}$ given $\{\bg,\bm\ell,\ell^{(2)}\}$, it follows that $\wh P = \psi^{\star} P$, $\wh n=\psi^{\star} n$ and $\wh{n}^{(2)} = \psi^{\star} n^{(2)}$. Thus, the causal character of $\mc D$ is the same as the one of $\psi^{\star}\mc D$, and in particular if $\mc D$ is null, so it is $\psi^{\star}\mc D$. The previous definition can be extended to the context of double null data as follows.

\begin{defi}
	\label{pullback2}
	Let $\{\mc D,\mc{\ul D},\mu\}$ be double null data and $\psi:\mc{\wh H}\to \mc H$, $\ul\psi:\mc{\ul{\wh H}}\to\mc{\ul H}$ diffeomorphisms. The pull-back double null data $\Xi^{\star}\{\mc D,\mc{\ul D},\mu\}$ is defined as $$\Xi^{\star}\{\mc D,\mc{\ul D},\mu\} \d \left\{\psi^{\star}\mc D,\ul\psi^{\star}\mc{\ul D},\ul\psi|_{\mc{\ul S}}^{\star}(\mu)\right\},$$ where $\psi^{\star}\mc D$ and $\ul\psi^{\star}\mc{\ul D}$ are the pull-backs in the sense of Def. \ref{pullback1}. 
\end{defi}

Since $\psi$ and $\ul\psi$ are diffeomorphisms, they preserve the boundaries $\mc{S}$ and $\mc{\ul S}$, and thus the map $\wh\phi\d\ul\psi\circ \phi\circ \psi^{-1}: \mc{\wh S}\to \mc{\ul{\wh S}}$ is a diffeomorphism. Then, $\Xi^{\star}\{\mc D,\mc{\ul D},\mu\}$ is still double null data, since it satisfies Def. \ref{defi_DND} with $\wh\phi$ and $\wh\mu \d \ul\psi|_{\mc{\ul S}}^{\star}(\mu)$. In the following proposition we find the necessary conditions for two DND to define two isometric Lorentzian manifolds.

\begin{prop}
	\label{prop_necesary_iso}
Let $\{\mc D,\ul{\mc D},\mu\}$ and $\{\mc{\wh{D}},\mc{\ul{\wh D}},\wh \mu\}$ be double null data satisfying the constraint equations \eqref{constraintsL} and let $(\mc M,g)$ and $(\mc{\wh M},\wh g)$ be respective developments. Suppose that there exists an isometry $\varphi:\mc M\to \mc{\wh M}$. Then there exist gauge parameters $(z,\zeta)$ and $(\ul z,\ul\zeta)$ in $\mc D$ and $\mc{\ul D}$, respectively, and a map $\Xi^{\star}$ as in Def. \ref{pullback2} such that $$\Xi^{\star} \{\mc{\wh{D}},\mc{\ul{\wh D}},\wh \mu\} = \mc G \left(\left\{\mc D,\ul{\mc D},\mu\right\}\right).$$
\begin{proof}
Let $\wh x^{\mu} = \{\wh u,\wh{\ul u},\wh x^A\}$ be coordinates on $\mc{\wh M}$ whose restrictions on $\mc{\wh H}$ and $\ul{\mc{\wh H}}$ satisfy \eqref{hola}. Define the coordinates $x^{\mu}$ on $\mc M$ by $x^{\mu}\d\wh x^{\mu}\circ \varphi$. Let $\Phi,\ul\Phi$ and $\xi,\ul\xi$ the embeddings and the riggings of $\{\mc D,\ul{\mc D},\mu\}$ in $(\mc M,g)$, and $\wh\Phi,\ul{\wh\Phi}$ and $\wh\xi,\ul{\wh \xi}$ be the embeddings and the riggings of $\{\mc{\wh{D}},\mc{\ul{\wh D}},\wh \mu\}$ in $(\mc{\wh M},\wh g)$. Since $(\varphi^{\star}\wh\xi) (u) = \wh\xi(u\circ\varphi^{-1}) = \wh\xi(\wh u)\neq 0$, there exist $z\in\mc F^{\star}(\mc H)$ and $\zeta\in\X(\mc H)$ such that $\varphi^{\star}\wh\xi = z(\xi+\Phi_{\star}\zeta)$ along $\Phi(\mc H)$. Since $\Phi(\mc H)$ is diffeomorphic to $\wh\Phi(\wh{\mc H})$ via $\varphi$ and both $\Phi$ and $\wh\Phi$ are embeddings, there exists a diffeomorphism $\psi$ making the following diagram commutative
\begin{center}
\begin{tikzcd}
	\mc H \arrow[r,"\psi"]\arrow[d,"\Phi"] & \wh{\mc H}\arrow[d,"\wh\Phi"]\\
	\mc M \arrow[r,"\varphi"] & \wh{\mc M}
\end{tikzcd}
\end{center}
Then $$\psi^{\star} \wh\bg = \psi^{\star} \wh\Phi^{\star} \wh g = \Phi^{\star}\varphi^{\star} \wh g = \Phi^{\star} g = \bg,$$ and $$\psi^{\star} \wh{\bm\ell} = \psi^{\star} \wh \Phi^{\star} \big(\wh g(\wh\xi,\cdot)\big) = \Phi^{\star} \varphi^{\star} \big(\wh g(\wh\xi,\cdot)\big) =\Phi^{\star}\left( g(z(\xi+\Phi_{\star}\zeta),\cdot)\right) = z (\bm\ell + \bg(\zeta,\cdot)).$$ Concerning $\wh\ell^{(2)}$,
\begin{align*}
\psi^{\star}\wh\ell^{(2)} &= \psi^{\star} \wh \Phi^{\star} \big(\wh g(\wh\xi,\wh\xi)\big)\\
& = \Phi^{\star}\left(z^2 g(\xi,\xi) + 2 z^2 g(\xi,\Phi_{\star}\zeta) + z^2 g(\Phi_{\star}\zeta,\Phi_{\star}\zeta)\right) \\
&= z^2(\ell^{(2)} + 2\bm\ell(\zeta)+\bg(\zeta,\zeta)).
\end{align*}
Finally, $$\psi^{\star} \wh\bY =\dfrac{1}{2} \psi^{\star} \wh \Phi^{\star} \lie_{\wh\xi}\wh g = \dfrac{1}{2}\Phi^{\star} \varphi^{\star}\lie_{\wh\xi}\wh g = \dfrac{1}{2}\Phi^{\star} \left(\lie_{z(\xi+\Phi_{\star}\zeta)} g \right)= z\bY + d z\otimes_s \bm\ell + \dfrac{1}{2}\lie_{z\zeta}\bg,$$ where the third equality holds because $\varphi^{\star}\wh\xi = z\left(\xi+\Phi_{\star}\zeta\right)$, $\varphi^{\star}\wh g=g$ and\footnote{This follows from the known formula $\varphi^{\star}\left(\lie_{\varphi_{\star}X} T\right) = \lie_{X} \left(\varphi^{\star}T\right)$ valid for any diffeomorphism $\varphi$, vector $X$ and tensor $T$ particularized to $T=\wh g$ and $\varphi_{\star}X = \wh\xi\Longrightarrow X = \varphi^{\star}\wh\xi = z\left(\xi+\Phi_{\star}\zeta\right)$.} $\varphi^{\star}\big(\lie_{\wh\xi} \wh g \big) = \lie_{\varphi^{\star}\wh\xi} \big(\varphi^{\star} g\big)$. Thus, recalling Def. \ref{defi_embedded}, $\psi^{\star} \wh{\mc D} = \mc G_{(z,\zeta)}(\mc D)$. The same argument on $\mc{\ul H}$ proves that there exist gauge parameters $(\ul z,\ul\zeta)$ on $\mc{\ul D}$ such that $\ul\psi^{\star} \wh{\mc{\ul D}} = \mc G_{(\ul z,\ul\zeta)}(\mc{\ul D})$. Finally, taking into account item 2. of Def \ref{def_embDND}, $$\ul\psi|_{\mc{\ul S}}^{\star}(\wh\mu) = \Phi|_{\mc{S}}^{\star}\varphi^{\star}\left(\wh g(\wh \nu,\wh{\ul\nu})\right) =\Phi|_{\mc{S}}^{\star}\left( g(z^{-1}\nu,\ul z^{-1}\ul\nu)\right) = z^{-1}\ul z^{-1} \mu.$$ Comparing with \eqref{transmu}, the result follows.
\end{proof}
\end{prop}

The previous proposition motivates defining the notion of isometric double null data as follows.

\begin{defi}
	\label{defi_isometricDND}
We say that two double null data $\{\mc D,\ul{\mc D},\mu\}$ and $\{\mc{\wh{D}},\mc{\ul{\wh D}},\wh \mu\}$ are isometric if there exist diffeomorphisms $\psi:\mc H\to\mc{\wh H}$ and $\ul\psi:\ul{\mc H}\to \wh{\ul{\mc H}}$ and gauge parameters $(z,\zeta)$ and $(\ul z,\ul\zeta)$ in $\mc D$ and $\mc{\ul D}$, respectively, such that the pull-back double null data $\Xi^{\star} \{\mc{\wh{D}},\mc{\ul{\wh D}},\wh \mu\}$ satisfies $$\Xi^{\star} \{\mc{\wh{D}},\mc{\ul{\wh D}},\wh \mu\} = \mc G \left(\left\{\mc D,\ul{\mc D},\mu\right\}\right).$$
\end{defi}

We conclude this paper by proving that the necessary conditions of Prop. \ref{prop_necesary_iso} are also sufficient. This result gives a geometric uniqueness statement of the characteristic problem of the Einstein field equations. Indeed, two isometric initial data are indistinguishable from a geometric point of view and thus they should have ``the same'' developments. The precise statement of the notion of uniqueness is given in the following theorem.

\begin{teo}
	\label{theorem_isometry}
Let $\{\mc D,\ul{\mc D},\mu\}$, $\{\mc{\wh{D}},\mc{\ul{\wh D}},\wh \mu\}$ be two isometric DND in the sense of Def. \ref{defi_isometricDND} with diffeomorphisms $\psi:\mc H \to \wh{\mc H}$ and $\ul\psi:\mc{\ul H}\to\wh{\mc{\ul H}}$ and satisfying the abstract constraint equations \eqref{constraintsL}, and let $(\mc M,g)$, $(\mc{\wh M},\wh g)$ be respective developments. Then there exist neighbourhoods $\mc U\subseteq\mc M$ and $\wh{\mc U}\subseteq \mc{\wh M}$ of $\mc H\cup\mc{\ul H}$ and $\mc{\wh H}\cup \mc{\ul{\wh H}}$, respectively, and a diffeomorphism $\varphi: \mc U\to\mc{\wh{\mc U}}$ such that $\varphi^{\star}\wh g = g$.
	\begin{proof}
We start by writing $\{\mc{\wh{D}},\mc{\ul{\wh D}},\wh \mu\}$ in the gauge of Lemma \ref{lema_compHG} w.r.t some coordinates $\{\wh x^{\ul a}\}=\{\wh u,\wh x^A\}$ and $\{\wh{\ul x}^{\ul a}\}=\{\wh{\ul u},\wh{\ul x}{}^A\}$ in $\mc{\wh H}$ and $\mc{\ul{\wh H}}$ satisfying \eqref{hola}, and $\{\mc D,\ul{\mc D},\mu\}$ in the gauge in which $\Xi^{\star} \{\mc{\wh{D}},\mc{\ul{\wh D}},\wh \mu\} =\left\{\mc D,\ul{\mc D},\mu\right\}$ holds. We want to show that $\Xi^{\star} \{\mc{\wh{D}},\mc{\ul{\wh D}},\wh \mu\}$ is written in the gauge of Lemma \ref{lema_compHG} w.r.t the coordinates $\{x^{\ul a}\}\d\{\wh x^{\ul a}\circ\psi\}$ and $\{\ul x^{\ul a}\}\d\{\wh{\ul x}^{\ul a}\circ\ul\psi\}$. Let $\wh V$ be the vector field defined in Thm. \ref{teo_HG} w.r.t $\{\wh x^{\ul a}\}$. First we prove that $\psi^{\star}\wh{V}$ is again the vector of Theorem \ref{teo_HG} but w.r.t $\{\psi^{\star} \wh x^{\ul a}\}$ (and analogously on $\ul{\mc H}$). Let $\{\wh{e}_c\}$ be a (local) basis of $\X(\wh{\mc H})$ and $\{e_c\d\psi^{\star}\wh{e}_c\}$ a (local) basis of $\X(\mc H)$. Since $\psi^{\star}\left(\wh{e}_c(\wh{x}^{\ul a})\right)=e_c(x^{\ul a})$, it follows $\psi^{\star}\wh{B}^c{}_{\ul a} = B^c{}_{\ul a}$. Moreover, since $\psi^{\star}\wh{\ol\nabla} = \ol\nabla$ (because $\Xi^{\star} \{\mc{\wh{D}},\mc{\ul{\wh D}},\wh \mu\} =\left\{\mc D,\ul{\mc D},\mu\right\}$), the pull-back of the Hessian of a function is the Hessian of the pullback of that function, and since $\psi^{\star}\wh{P}=P$, it turns out that $\psi^{\star}\big(\wh{P}^{ab}\wh{\ol\nabla}_a\wh{\ol\nabla}_b\wh{x}^{\ul a}\big) = P^{ab}\ol\nabla_a\ol\nabla_b x^{\ul a}$, and thus $\psi^{\star}\wh{V}^c = \psi^{\star}\big(\wh{B}^c{}_{\ul a} \wh{\square}_{\wh{P}} \wh{x}^{\ul a}\big)= B^c{}_{\ul a}\square_P x^{\ul a}$. Therefore $\Xi^{\star} \{\mc{\wh{D}},\mc{\ul{\wh D}},\wh \mu\}$ is written in a harmonic gauge w.r.t $\{x^{\ul a}\}$ and $\{\ul x^{\ul a}\}$. Moreover, since $\psi^{\star}\wh\ell^{(2)} =\ell^{(2)} =  0$, $\psi^{\star} \wh{\bm\ell}_{\para}= \bm\ell_{\para}= 0$ on $\mc{S}$, $\ul\psi^{\star}\wh{\ul\ell}{}^{(2)}=\ul\ell^{(2)}=0$, $\ul\psi^{\star} \wh{\bm{\ul\ell}}_{\para}=\bm{\ul\ell}_{\para}=0$ on $\mc{\ul S}$, as well as $\ul\psi_{\ul{\mc S}}^{\star}(\wh\mu) = \ul\psi_{\ul{\mc S}}^{\star}\left(\wh{\ul n}({\wh u})\right)=\left(\ul\psi^{\star}\wh{\ul n}\right)(u)$ on $\mc{\ul S}$ and similarly $\ul\psi|_{\ul{\mc S}}^{\star}(\wh\mu) = \left(\psi^{\star}\wh n\right)(\ul u)$ on $\mc{S}$ (we omit the $\phi^{\star}$ for simplicity), we conclude that the data $\Xi^{\star} \{\mc{\wh{D}},\mc{\ul{\wh D}},\wh \mu\}$ is written in the gauge of Lemma \ref{lema_compHG} w.r.t the coordinates $\{x^{\ul a}\}$ and $\{\ul x^{\ul a}\}$.\\

\begin{figure}
	\psfrag{a}{$(\mc M,g)$}
	\psfrag{b}{$(\wh{\mc M},\wh g)$}
	\psfrag{e}{$\varphi$}
	\psfrag{c}{$\mc U$}
	\psfrag{d}{$\mc{\wh U}$}
	\centering
	\includegraphics[width=0.8\linewidth]{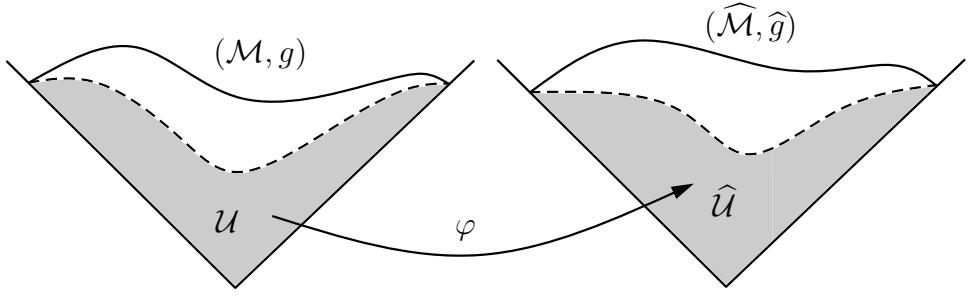}
	\caption{Given isometric DND, there exist isometric neighbourhoods $\mc U$ and $\mc{\wh U}$ of the initial data.}
	\label{fig:drawing-1}
\end{figure}

Let $(\mc M,g)$ be a development of $\{\mc D,\ul{\mc D},\mu\}$ with embeddings $\Phi$, $\ul \Phi$ and riggings $\xi$, $\ul\xi$ and let $(\mc{\wh M},\wh g)$ be the same but everything with a ``\ $\wh{}$\ ''. Let $\{x^{\mu}\}$ be the harmonic coordinates on $(\mc M,g)$ restricting to the given ones at $\mc H\cup\mc{\ul H}$ and satisfying $\Phi(\mc H)=\{u=0\}$, $\ul \Phi(\mc{\ul H})=\{\ul u=0\}$ (and the same with ``\ $\wh{}$\ ''). It is straightforward to see that the rigging vectors are given by $\xi=\partial_u$, $\ul\xi=\partial_{\ul u}$ and $\wh\xi=\partial_{\wh u}$, $\wh{\ul \xi}=\partial_{\ul{\wh u}}$ (we refer the reader to the proof of Theorem 7.15 of \cite{Mio1} for the details). Choosing suitable neighbourhoods $\mc U\subseteq\mc M$ of $\mc H\cup\mc{\ul H}$ and $\wh{\mc U}\subseteq \mc{\wh M}$ of $\mc{\wh H}\cup\mc{\wh{\ul H}}$ (see Figure \ref{fig:drawing-1}), we can define the diffeomorphism $\varphi$ by $x^{\mu} = \wh{x}^{\mu}\circ\varphi$, which by construction restricts to the given diffeomorphisms $\psi:\mc H\to\mc{\wh H}$ and $\ul\psi:\ul{\mc H}\to \wh{\ul{\mc H}}$. Since $(\mc{\wh M},\wh g)$ is a solution of the EFE with cosmological constant $\Lambda$, so it is $(\mc M,\varphi^{\star}\wh g)$. Moreover, $\square_{\varphi^{\star}\wh g} x^{\mu} = 0$, so both $g$ and $\varphi^{\star}\wh g$ are a solution of the reduced equations in the coordinates $\{x^{\mu}\}$. By theorem 1 of \cite{Rendall}, in order to prove that $g=\varphi^{\star}\wh g$ on $\mc U$, we only need to show that their restrictions to $\mc H\cup\ul{\mc H}$ agree, which follows at once after recalling $\Xi^{\star} \{\mc{\wh{D}},\mc{\ul{\wh D}},\wh \mu\} =\left\{\mc D,\ul{\mc D},\mu\right\}$ and using that $\varphi$ restricts to $\psi$ and $\ul\psi$ on $\mc H$ and $\mc{\ul H}$, respectively.
\end{proof}
\end{teo}
% $(\wt\varphi^{\star}\wh g)_{u\, u}|_{\mc H} = \varphi^{\star}\left(\wh g(\wh\xi,\wh\xi)|_{\mc{\wh H}}\right) = \varphi^{\star}\wh\ell^{(2)} = \ell^{(2)} = g_{u\, u}|_{\mc H}.$

\begin{rmk}
	
	The  geometric existence and uniqueness statements in Theorems \ref{main} and \ref{theorem_isometry} above refer to a neighbourhood of the intersection surface. This limitation arises because we rely on Rendall's theorem to establish the results. In \cite[Theorem A.1]{rodnianski} the authors prove the existence of the spacetime in a neighbourhood of the full initial hypersurfaces. We expect that a suitable application of this result should allow us to extend our geometric existence and uniqueness results to the full double null data as well. This is however not immediate because the field variables and gauge fixing conditions are different. We intend to analyze this issue in  a future work.
	
\end{rmk}

\section{Conclusions}

In this paper we have extended our previous definition of \textit{double null data} to a gauge-covariant one by introducing the notion of normal pair, allowing us to encode abstractly the extrinsic geometric information of the intersection surface. This leads to the so-called compatibility conditions, which turns out to be necessary and sufficient for any double null data to be embeddable in some Lorentzian manifold. Moreover, if the abstract constraint equations are fulfilled, there exists a $\Lambda$-vacuum spacetime in which the double null data can be embedded. Finally, we introduced the notion of \textit{isometric double null data} and prove that the developments of two isometric double null data satisfying the abstract constraints are isometric, achieving a geometric uniqueness result for the characteristic problem. The results of this paper and our previous one \cite{Mio1} provide a satisfactory geometrization of the characteristic Cauchy problem for the Einstein equations and put it on the same footing as the standard Cauchy problem.

\begin{appendices}

\section{Abstract constraint tensor in a section}
\label{appendix}

Let $\mc D=\{\mc H,\bg,\bm\ell,\ell^{(2)},\bY\}$ be null hypersurface data admitting a section $i:\mc S\hookrightarrow \mc H$. As shown in \cite[Cor. 4.4]{Mio1}, one can always choose a gauge in which $\bm\ell_{\para}=0$ and $\ell^{(2)}=0$ on $\mc S$. In this appendix we compute the pullback of the abstract constraint tensor $\bm{\mc R}$ as defined in \eqref{Rab} into $\mc S$ in the aforementioned gauge.

\begin{lema}
	\label{lemapendice}
Let $\mc D=\{\mc H,\bg,\bm\ell,\ell^{(2)},\bY\}$ be null hypersurface data admitting a section $i:\mc S\hookrightarrow \mc H$ and let $\bm{\mc R}$ be the abstract constraint tensor \eqref{Rab}. Then the pullback of $\bm{\mc R}$ into $\mc S$ in a gauge in which $\ell^{(2)}=0$ and $\bm\ell_{\para}=0$ on $\mc S$ takes the following form
\begin{equation}
	\label{RAB}
	\begin{aligned}
		\bm{\mc R}_{AB} & = R^h_{AB} + \left(2\bY(n,n) - \tr_h\bK\right) \bY_{AB} + \left(n\left(\ell^{(2)}\right) - \tr_h\bY\right)\bK_{AB}\\
		&\quad + 4h^{CD}\bK_{C(A}\bY_{B)D} + 2\nabla_{(A}^h\bm\tau_{B)} + 2\nabla_{(A}^h\left(\lie_n\bm\ell\right)_{B)}- 2\left(\lie_n\bPi\right)_{(AB)}- 2\bm\tau_A\bm\tau_B.
	\end{aligned}
\end{equation}
\begin{proof}
Let $A$ and $B$ be the tensors defined by 
\begin{align}
	A_{bca}&\d\bm\ell_d\ol{R}^d{}_{bca}+2\ell^{(2)}\ol{\nabla}_{[a} \bK_{c]b} +\bK_{b[c}\ol\nabla_{a]} \ell^{(2)},\label{A} \\
	B_{abcd}&\d\bg_{af}\ol{R}^f{}_{bcd}+2\ol\nabla_{[d}\left(\bK_{c]b}\bm\ell_a\right)+2\ell^{(2)}\bK_{b[c}\bK_{d]a}.\label{B}
\end{align}
Comparing \eqref{Rab} with \eqref{A}-\eqref{B}, the tensor $\bm{\mc R}$ can be written as 
\begin{equation}
	\label{Rab2}
	\bm{\mc R}_{ab} = B_{acbd}P^{cd} + \left(A_{bca}+A_{acb}\right) n^c.
\end{equation}
Let $\{e_A\}$ be a (local) basis of $T\mc S$ with dual basis $\{\theta^A\}$, i.e., $\theta^A(e_B)=\delta^A_B$. Then $\{n,e_A\}$ is a (local) basis of $T\mc S\oplus\mbox{span}\{n\}$, and since $\bm\ell_{\para}=0$, $\{\bm\ell,\theta^A\}$ is the dual basis of $\{n,e_A\}$. By equations \eqref{Pell}-\eqref{Pgamma}, the tensor $P$ takes the following form in the basis $\{n,e_A\}$
\begin{equation}
	\label{P}
	P = h^{AB} e_A\otimes e_B.
\end{equation} 
In order to write down the tensor $\bm{\mc R}_{AB}\d \bm{\mc R}_{ab} e^a_A e^b_B$ we divide the computation into two parts, namely $B_{acbd}P^{cd}e^a_A e^b_B$ and $A_{bca} n^ce^a_A e^b_B$. To calculate the former we need to recall some results concerning the relation between $\ol\nabla$ and the Levi-Civita connection $\nabla^h$ of $h\d i^{\star}\bg$. As a consequence of Prop. 3.5 and equation (30) of \cite{Mio1} together with $\bm\ell_{\para}=0$ and $\ell^{(2)}=0$, the relation between $\ol\nabla$ and $\nabla^h$ is
\begin{equation}
	\label{connectiondeco}
	\ol\nabla_X Y = \ol\nabla^{h}_XY - \bY(X,Y)n,\qquad X,Y\in\X(\mc S).
\end{equation}
As usual, this decomposition allows us to relate the completely tangential components of the curvature tensor of $\ol\nabla$ with the curvature tensor of $\nabla^{h}$ via a Gauss-type identity. The explicit expression is obtained in \cite[Prop. 3.7]{Mio1} and in the present gauge reads
\begin{equation}
	\label{gauss2}
	\bg\left(W,\ol R(X,Y)Z\right) = h\left(W, R^{h}(X,Y)Z\right) - \bY(Y,Z)\bK(X,W) + \bY(X,Z)\bK(Y,W),
\end{equation}
where $\ol{R}$ and $R^{h}$ are the curvature tensors of $\ol\nabla$ and $\nabla^h$, respectively. We now have all the ingredients needed to compute $B_{acbd}P^{cd}e^a_A e^b_B$. Taking into account that $\ell^{(2)}=0$ and using \eqref{P}, 
\begin{equation}
	\label{appaux1}
	B_{acbd}P^{cd}e^a_A e^b_B = \left(\bg_{cf}\ol{R}^f{}_{adb} + 2\ol\nabla_{[b}\left(\bK_{d]a}\bm\ell_c\right)\right)h^{CD}e_C^ce_D^de^a_A e^b_B.
\end{equation}
For the first term we employ Gauss identity \eqref{gauss2} and equation \eqref{P},
\begin{align*}
\bg_{cf}\ol{R}^f{}_{adb}h^{CD}e_C^ce_D^de^a_A e^b_B & = \bg\left(e_C,\ol R(e_D,e_B)e_A\right) h^{CD} \\
&= R^h_{AB} - \bY_{AB}\tr_h\bK + h^{CD}\bY_{DA}\bK_{BC} ,
\end{align*}
where $R^h_{AB}$ is the Ricci tensor of $h$. In this appendix, when no confusion arises we denote with the same symbol a tensor on $\mc H$ and its pullback on $\mc S$. For the second term of \eqref{appaux1} we use \eqref{olnablaell}, which in this gauge is $\ol\nabla_a\bm\ell_b \st{\mc S}{=} \bPi_{ab}$, and the fact that $\bPi_{AB} \st{\mc S}{=} \bY_{AB}$ (because $\bm\ell_{\para}=0$ and thus $2i^{\star}\bF= i^{\star} d\bm\ell = d\bm\ell_{\para}=0$). Then, $$2h^{CD}e_C^ce_D^de^a_A e^b_B\ol\nabla_{[b}\left(\bK_{d]a}\bm\ell_c\right) = 2h^{CD}e_C^ce_D^de^a_A e^b_B \bK_{a[d} \ol\nabla_{b]}\bm\ell_c = -2 h^{CD}\bK_{A[B}\bY_{D]C},$$ where the first equality holds because $\bm\ell_{\para}=0$. Combining the two terms, \eqref{appaux1} finally reads
\begin{equation}
	\label{appaux0}
B_{acbd}h^{CD}e_C^ce_D^de^a_A e^b_B = R^h_{AB} - \bY_{AB}\tr_h\bK  - \bK_{AB}\tr_h\bY+ 2h^{CD}\bY_{C(A}\bK_{B)D} .
\end{equation}
Next we compute the terms of the form $A_{bca} n^ce^a_A e^b_B$. In the present gauge the quantity $A_{bca} n^c$ is simply (note that $\ell^{(2)}$ is not assumed to be zero off $\mc S$)
\begin{equation}
	\label{secondterm}
	n^c\left(\bm\ell_d\ol{R}^d{}_{bca}+\bK_{b[c}\ol\nabla_{a]} \ell^{(2)}\right)=n^c\bm\ell_d\ol{R}^d{}_{bca}-\dfrac{1}{2}\bK_{ba}n\left( \ell^{(2)}\right),
\end{equation}
where $\bK(n,\cdot)=0$ has been used. The first term in \eqref{secondterm} can be computed directly from the Ricci identity,
\begin{align}
	n^c\bm\ell_d\ol{R}^d{}_{bca} & = 2n^c\ol\nabla_{[a}\ol\nabla_{c]}\bm\ell_b \nonumber\\
	&= 2n^c \left(\ol\nabla_{[a}\bPi_{c]b} - \bK_{b[c}\ol\nabla_{a]}\ell^{(2)}\right)\nonumber\\
	&=n^c\ol\nabla_a\bPi_{cb} -\ol\nabla_n\bPi_{ab} + \bK_{ab}n\left(\ell^{(2)}\right),\label{RicciId}
\end{align}
where we used $\ol\nabla_a\bm\ell_b = \bPi_{ab} - \ell^{(2)}\bK_{ab}$ and that $\bK(n,\cdot)=0$. Contracting the first term in \eqref{RicciId} with $e^a_Ae^b_B$ and using \eqref{olnablannull} and \eqref{cartan}
\begin{align*}
e^a_Ae^b_B n^c\ol\nabla_a\bPi_{cb} & = e^a_Ae^b_B\ol\nabla_a\left(\bPi_{cb}n^c\right) - e^a_Ae^b_B\bPi_{cb}\ol\nabla_a n^c\\
&=e^a_Ae^b_B\ol\nabla_a\left(\bPi_{bc}n^c   +\lie_n\bm\ell_b \right)  - e^a_Ae^b_B\bPi_{cb}\left(P^{cd}\bK_{da} - \bPi_{ad}n^cn^d\right)\\
&=e_A\left(\bPi(e_B,n) +(\lie_n\bm\ell)(e_B)\right) - \left(\bPi(\cdot,n)+\lie_n\bm\ell\right)\left(\ol\nabla_{e_A} e_B^b\right)\\
&\quad - h^{CD}e_C^ce_D^de^a_Ae^b_B \bPi_{cb}\bK_{da} + e^a_Ae^b_B \left(\bPi_{bc} + \lie_n\bm\ell_b\right)\bPi_{ad}n^cn^d.
%&= \left(\nabla_{e_A}^h\left(\bm\tau + \lie_n\bm\ell\right)\right)(e_B) + \bY(n,n)\bY(e_A,e_B) - h^{CD}\bY_{CB}\bK_{DA}\\
%&\quad + \bm\tau_A\bm\tau_B + \bm\tau_A \left(\lie_n\bm\ell\right)_B
\end{align*}
Introducing \eqref{connectiondeco} and recalling $\bPi(n,n)=\bY(n,n)$, $\bPi_{AB}=\bY_{AB}$ and the fact that in this gauge $\bPi(e_A,n)=\bm\tau_A$ (see Def. \ref{defi_foliationtensors}) this expression finally yields
\begin{equation}
	\label{appaux2}
	\begin{aligned}
e^a_Ae^b_B n^c\ol\nabla_a\bPi_{cb} &= \left(\nabla_{e_A}^h\left(\bm\tau + \lie_n\bm\ell\right)\right)(e_B) + \bY(n,n)\bY_{AB} \\
&\quad- h^{CD}\bY_{CB}\bK_{DA}+ \bm\tau_A\bm\tau_B + \bm\tau_A \left(\lie_n\bm\ell\right)_B ,
\end{aligned}
\end{equation}
where we also used $\left(\lie_n\bm\ell\right)(n)=\lie_n\left(\bm\ell(n)\right)=0$. Equation \eqref{olnablannull} can be rewritten as 
\begin{equation}
\label{olnablannull2}
\ol\nabla_X n = K^{\sharp}(X) - \bPi(X,n)n,
\end{equation}
where $K^{\sharp}$ is the endomorphism defined by $K^{\sharp}(X)\d P\left(\bK(X,\cdot),\cdot\right)$, or in abstract index notation $(K^{\sharp})^a{}_b = P^{ac}\bK_{cb}$. Using $\ol\nabla_n X = \ol\nabla_X n + \lie_n X$ together with \eqref{olnablannull2}, the contraction of the second term in \eqref{RicciId} with tangential directions can be written as
\begin{align*}
	X^a Y^b \ol\nabla_n\bPi_{ab} & = n\left(\bPi(X,Y)\right) - \bPi\left(\ol\nabla_n X, Y\right) - \bPi\left(X,\ol\nabla_n Y\right)\\
	&=\lie_n \left(\bPi(X,Y)\right) - \bPi\left(\ol\nabla_{X} n + \lie_n X, Y\right) - \bPi\left(X,\ol\nabla_{Y} n +\lie_n Y\right)\\
	&= \left(\lie_n\bPi\right)(X,Y) - \bPi\left(K^{\sharp}(X)-\bPi(X,n)n, Y\right) - \bPi\left(X,K^{\sharp}(Y) - \bPi(Y,n)n\right),
\end{align*}
%or in abstract index notation,
%\begin{equation}
%	\label{appaux3}
%	\ol\nabla_n\bPi_{ab} = \lie_n\bPi_{ab} - P^{cd}\bPi_{cb}\bK_{da} + \bPi_{ac}\bPi_{db}n^cn^d - P^{cd}\bPi_{ac}\bK_{db} + \bPi_{bc}\bPi_{ad}n^cn^d.
%\end{equation}
and therefore
\begin{equation}
	\label{appaux3}
e^a_Ae^b_B \ol\nabla_n\bPi_{ab} = \left(\lie_n\bPi\right)_{AB} - h^{CD}\bY_{CB}\bK_{DA} + 2\bm\tau_A\bm\tau_B + \bm\tau_A \left(\lie_n\bm\ell\right)_B - h^{CD}\bY_{AC}\bK_{DB} .
\end{equation}
Contracting \eqref{RicciId} with $e^a_Ae^b_B$ and introducing \eqref{appaux2} and \eqref{appaux3},
\begin{align*}
e^a_Ae^b_Bn^c\bm\ell_d\ol{R}^d{}_{bca} &= \left(\nabla_{e_A}^h\left(\bm\tau + \lie_n\bm\ell\right)\right)(e_B) -\left(\lie_n\bPi\right)_{AB}   + \bY(n,n)\bY_{AB}    \\
&\quad + h^{CD}\bY_{AC}\bK_{DB}- \bm\tau_A\bm\tau_B  + n\left(\ell^{(2)}\right)\bK_{AB},
\end{align*}
and thus
\begin{equation}
	\label{appaux5}
	\begin{aligned}
		\left(A_{acb}+A_{bca}\right)n^ce_A^ae_B^b & = 2\nabla_{(A}^h\bm\tau_{B)} + 2\nabla_{(A}^h\left(\lie_n\bm\ell\right)_{B)} - 2\left(\lie_n\bPi\right)_{(AB)} + 2\bY(n,n)\bY_{AB}  \\
		&\quad + 2h^{CD}\bK_{C(A}\bY_{B)D} - 2\bm\tau_A\bm\tau_B + n\left(\ell^{(2)}\right)\bK_{AB},
	\end{aligned}
\end{equation}
Finally, combining \eqref{appaux0} and \eqref{appaux5}, \eqref{RAB} follows.
\end{proof}
\end{lema}

\end{appendices}

\section*{Acknowledgements}
This work has been supported by Projects PID2021-122938NB-I00 (Spanish Ministerio de Ciencia e Innovación and FEDER ``A way of making Europe'') and SA096P20 (JCyL). G. Sánchez-Pérez also acknowledges support of the PhD. grant FPU20/03751 from Spanish Ministerio de Universidades. We are very grateful to Miguel Manzano for useful comments.

\begingroup
\let\itshape\upshape

\renewcommand{\bibname}{References}
\bibliographystyle{acm}
\bibliography{biblio}

\end{document}